# General problems of the internal gravity waves linear theory

**Vitaly V. Bulatov and Yuriy V. Vladimirov**

Institute for Problems in Mechanics
Russian Academy of Sciences
Pr.Vernadskogo 101-1, 117526 Moscow, Russia
**bulatov@index-xx.ru**

**Abstract**
The internal gravity waves are the oscillations present in the gravitational field of the stratified medium, that is the mediums which density raises with the depth change. If the equilibrium state of the component volume of this medium is disturbed, for example, upward, then it will become more heavy, than the medium surrounding it, and Archimedian forces will cause its motion back to its equilibrium position. The main parameter of any oscillation system is the oscillation frequency, and it is determined by the ratio of two factors - the restoring forces seeking to return the disturbed system to its equilibrium position and the inertial forces. For the internal waves the restoring forces are proportional to the vertical gradient of density of the liquid, and the inertial forces are proportional to the density itself. The oscillating frequency typical for the internal gravitational waves is the value $N(z) = \left[ -\frac{g}{\rho(z)} \frac{d\rho(z)}{dz} \right]^{\frac{1}{2}}$ called the buoyancy frequency or Brunt-Väisälä frequency. Here $\rho(z)$ - the density as the function of the depth $z$, $g$ - the acceleration of the gravity, the sign "-" originates due to the fact, that the density raises with the increasing depth and consequently $\frac{d\rho(z)}{dz} < 0$. The paper has considered the planar internal gravitational waves in the exponentially stratified medium, that is in the medium with the constant distribution of Brunt-Väisälä frequency in depth, analyzed the problem of the waves reflection from the planar boundaries, defined the Green's function for the equation of the internal gravity waves in the exponentially stratified medium of the of the endless depth, and also outlined the information on the main properties of the internal gravity waves in the stratified layer of the terminated depth and in the stratified mediums with average shift streams.

### 0. Introduction

Now in connection with the new problems arising in geophysics, oceanology, physics of atmosphere, usage of the cryogenic liquids in the engineering sphere, as well as the problems of protection and study of the medium, operation of the complex hydraulic engineering facilities, including the marine oil producing complexes, and a number of other actual problems facing the science and engineering we can observe the growth of interest to the research of the dynamics of the wave movements of the different inhomogeneous liquids and, in particular, the stratified liquids. This interest is caused not only by the practical needs, but also by the need to have the solid theoretical base to solve the arising problems. It is necessary to note, that solution of the problems of the mechanics of continua and hydrodynamics always served as the stimulus of new directions in mathematics and mathematical physics. As the illustration to the above may serve the stream of the new ideas in the theory of the nonlinear differential equations, and also the discovery of the startling dependencies between the can be appearing the different branches of mathematics, that has followed after exploration of Cartevega de Vriza equation for the waves on the shallow water. Certainly, for the detailed description of the big amount of the natural phenomena connected with the dynamics of the stratified heterogeneous in the horizontal direction and the non-stationary mediums, it is necessary to use the sufficiently developed mathematical models, which as a rule are the rather complex nonlinear multiparametric mathematical models and for their full-size research only the numerical methods are effective.

Certainly, for the detailed description of the big amount of the natural phenomena connected with the dynamics of the stratified heterogeneous in the horizontal direction and the non-stationary mediums, it is necessary to use as the base the sufficiently developed mathematical models, which as a rule are the rather complex nonlinear multiparametric mathematical models and for their full-size research only the numerical methods are effective. However in some cases the initial qualitative idea of the amount of the studied phenomena one can receive on the basis of use of the more simple asymptotical models and the analytical methods of their research. It becomes evident, that in this respect the problems of the dynamics of the internal gravitational waves in the heterogeneous mediums are rather indicative. Even at the use of the linear models their solutions are rather specific and determine the independent mathematical interest alongside with the nontrivial. physical corollaries.

The interest to the internal gravitational waves is caused by their wide presence in the nature. Both the air atmosphere, and the oceans are stratified. Reduction of the air pressure and its density at the increase of the elevation are well-known. But the sea water is also stratified. Here the raise of the water density with the increase of its depth is determined, mainly, not by the rather small compressibility of the water, but by the fact, that with the increase of the depth, as a rule, the temperature of the water is decreasing, and its saltiness grows. In the capacity of the stratified medium, as a rule, one consider s the medium, the physical characteristics (density, dynamic viscosity and others) of which in the medium stationary status are changing only along some concrete direction. Stratification of the natural mediums (the ocean, the atmosphere) can be caused by the different physical reasons, but the most often – by the gravity. This force creates in the stratified medium such a distribution of the particles of the dissolved in it salts and suspensions, at which it forms the heterogeneity of the medium along the direction of the gravitational field in the stratified medium.

This heterogeneity is called the density stratification. The stratification of density, as the experimental and natural observation show, renders the most essential influence, as compared with other kinds of stratification, on the dynamic properties of the medium and on the processes of distribution in the medium of the wave movements. Consequently at consideration of the wave generations in the stratified mediums usually neglect all other kinds of wave stratification, except for the density stratification, and in the capacity of the stratified medium they consider the medium with density stratification caused by the gravity.

In the real oceanic conditions the density changes are small, the periods of oscillations of the internal waves are changing from several minutes (in the layers with rather fast change of the temperatures and the depth) up to the several hours. Such great periods of the fluctuations means, that even at the big amplitude of the internal waves, but they can achieve dozens of meters along the vertical direction , the speeds of the particles in the internal wave are low – for the vertical components the speeds of the particles have the order of mm/s, and for the horizontal - cm/s. Therefore the dissipative losses – the losses caused by of the liquid viscosity in the internal waves are very small, and the waves propagation can propagate practically without fading within the big distances,. At that the speed of propagation of the internal waves in the ocean is low - the order of dozens of cm/s.

These properties of the internal gravitational waves mean, that they can keep the information about the sources of their generations for the long time. Unfortunately, it is very difficult to orientate in this information because the internal waves pass the dozens and hundreds of kilometers from the source the generations up to the place of supervision; and practically everywhere, where there is the stratification of the ocean takes place, we can observe the internal waves, but simultaneously we can "hear" the "voices" of the most different sources. At that the qualitative (and the quantitative) properties of the internal waves, caused by that or other concrete source depend not only on its physical nature, and also on its spatial and time distribution, but also depends on the properties of the medium located between the source of the waves and the place of the observation.

The internal waves represent the big interest not only from the point of view of their applications. They are of the interest to the theorists occupied with the problem of propagation ща the waves, as the internal waves properties in many respects differ from the properties of the accustomed to us the acoustic or electromagnetic waves. For example, for short harmonious internal waves of the following kind $A \exp(ikS(x, y, z) - i\omega t)$, where $k \gg 1$) – the beams are directed not perpendicularly to the wave fronts – to the surfaces of the equal phase $S = \text{const}$, but along these surfaces.

The stratification, or the layered structure of the natural mediums (oceans and the air atmosphere) causing formation of the internal gravitational waves plays then appreciable role in different oceanic and atmospheric processes and influences on the horizontal and vertical dynamic exchanges. The periods of the internal waves can make from several minutes up to several hours, the lengths of the waves can to achieve up to dozens of kilometers, and their amplitudes can exceed dozens of meters. The physical mechanism of formation of the internal waves is simple enough: if in the steadily stable stratified medium has appeared a generation, which has caused the particle out its balance state, then under action of gravity and the buoyancy the particle will make fluctuations about its balance position.

The theory of the wave movements of the stratified liquid being the section of the modern hydrodynamics is quickly developing recently and rather interesting in the theoretical aspect as well as it is connected with the major applications in the engineering field (hydraulic engineering, shipbuilding, navigation, energy) and in geophysics (oceanology, meteorology, hydrology, preservation of the environment). Now the majority of the applied problems, concerning the waves generation caused by various generations are solved just in the linear aspect, that is considering the assumption, that the amplitude of the wave movements is small in comparison with length of the wave. The relative simplicity of the solution of the linear equations as compared with the solution of the complete nonlinear problem, the modern development of the corresponding mathematical tools and the computer engineering allows to meet many challenges of practice.

Initially the theory of wave movements of the stratified liquid was developing as the theory of superficial waves describing the behavior of the free surface of the liquid being in the gravity field. Later it has been understood, that the superficial waves represent the special type of the waves existing on the border of the separation of the liquids of various densities, which in turn represent the special case of the internal waves in the medium heterogeneous (stratified) in density. In the real natural mediums the non-uniform distribution of density may take place both in the vertical, and in the horizontal directions. At that considering the existing heterogeneity of the medium both in the vertical the horizontal directions, and also its nonstationarity at research of the distribution of the internal gravitational waves require to use the special mathematical tools. The monograph as a rule considers only the case of the change of the density in the vertical direction. At that it is supposed, that the distribution of the density is steady, that is the density does not decrease with the change of the depth.

The reasons of initiation of the superficial and internal waves in the ocean and the atmosphere are very different: the fluctuations of the atmospheric pressure, the flow past of the bottom asperities, movement of the surface or the underwater ship, deformation in the density field, the turbulent spots formed by any reasons, the bottom shift or the underwater earthquake, the surface or underwater explosions, etc. One of the mechanisms of generation of the internal gravitational waves may be excitation of the wave fields caused by, for example, at movement (flow past) of the solid bodies, the turbulent spots, the water lenses and the other non-wave formations with the abnormal characteristics. At the solution of the problem of generation of the internal waves, for example, the stream of the stratified liquid, which is flowing past some non-local obstacle, the most widespread are the following two methods. The first method considers the numerical solution of the linearized system of the equations of the hydrodynamics describing the internal waves, to the shortcomings of which one may refer the limitation of the field of integration of the space, in which there is a possibility to have the numerical solution of the problem. The second method provides for substitution of the function describing the form of the non-local source of the generation either with the rather simple function, or with the system of the dot sources taken with a some weight. It is quite obvious, that at solution of the problem in such a way there is the problem of estimation of borders of applicability of use of the various simplifying approximations of the flow-past obstacle.

Generally the system of the linear equations describing the small movements of the originally quiescent incompressible non-viscous liquid in the system of the Cartesian coordinates $\mathbf{x} = (x, y, z)$ with the axis $z$ directed vertically upwards, looks like

$$\text{div } \mathbf{U} = Q(\mathbf{x}, t) \tag{0.1}$$

$$\rho_0 \frac{\partial \mathbf{U}}{\partial t} + \text{grad} p + F = S(x,t)$$

$$\frac{\partial \rho}{\partial t} + \rho_0' W = K(x,t)$$

Where $\mathbf{U} = (U_1, U_2, W)$, $p$, $\rho$ - generations of the vector of speed, pressure and density; $\rho_0(z)$ - density of the liquid in the quiescent state; $F=(0,0,g\rho)$, $g$ - acceleration of the gravity; the stroke means differentiation by a variable $z$. Functions $Q$, $S$, $K$ represent intensities of distributions of the sources of weight, pulses and density accordingly. Boundary conditions on the free surface - $(z=0)$ and on the flat bottom - $(z=-H)$ look like :
$W=\partial\eta/\partial t$, $p$-$g\rho_0\eta$=P(x,y,t) $(z=0)$, (0.2)

$W=Z(x,y,t)$ $(z=-H)$.

Here function $\eta(x, y, t)$ describes the vertical displacement of the free surface; P - external pressure, acting on the free surface; and Z – the vertical speed of the bottom. The initial conditions at $t=0$ are as follows:

$$U = U^*(x), \rho=\rho^*(\mathbf{x}), \eta=\eta^*(x, y) \tag{0.3}$$

Where functions $U^*(x)$, $\rho^*$, $\eta^*$ - initial values of generations of the vector of speed, density and elevation of the free surface. To ensure the correct performance of the condition it is required to meet the following condition:

$$\text{div } U^*(x) = Q \qquad (t=0).$$

By virtue of the linearity of the problem the forced waves are represent by the superposition of the free harmonious waves described by the homogeneous system (0.1) and the homogeneous boundary and initial conditions of (0.2), (0.3). The system (0.1) can be reduced to one equation for any of required functions, usually it is done for the vertical component of speed. At that the homogeneous system (0.1) and the homogeneous boundary conditions (0.2) may be presented in the form of :

$$\frac{\partial^2}{\partial t^2}\left((\frac{\partial^2}{\partial z^2}+\Delta)W - \frac{N^2(z)}{g}\frac{\partial W}{\partial z}\right) + N^2 \Delta W = 0$$

$$\frac{\partial^3 W}{\partial z \partial t^2} - g\Delta W = 0$$

(0.4)

(z=0), W=0 (z =-H),

Where $\Delta$ - is the flat (on variables *x, y*) Laplace operator; $N^2(z) = -g\,\rho_0'/\rho_0$ -. The equation (0.4) is to some extend simplified after introduction of Boissinesq approximation. At usage of this approximation in the equations of the pulse preservation (0.1) the difference of the density from some constant value $\rho_s$, is considered only in the member describing floatability, in the inertial members the real density is replaced with the value $\rho_s$, and the equation (0.4) is reduced to the kind:

$$\frac{\partial^2}{\partial t^2}(\frac{\partial^2}{\partial z^2}+\Delta)W + N^2 \Delta W = 0$$

(0.5)

The function N(z) (in Boissinesq approximation $N(z) = \sqrt{-g\rho_0'/\rho_s}$ ) is one of the basic characteristics of the stratified liquid, and has the fundamental value in the theory of the internal waves and is called the buoyancy frequency or Brunt-Väisälä frequency. The value T=2π/N defines Wjaysjal-Brent period. For the real ocean and the atmosphere the value T varies from minutes up to several hours, and for the stratified liquid produced in the laboratory, it can make some seconds.

Homogeneity of the equations (0.4), (0.5) and their boundary conditions at the variables x, y, t allow to look for the elementary wave solutions in the field of the plane waves:

$$W(\mathbf{x}, t) = \varphi(z)\, e^{i(\mathbf{k}\cdot\mathbf{r}-\omega\cdot t)}$$

Where **k** is the wave vector in the plane x, y; $\omega$ - oscillations frequency; **r** = (x, y).

For function $\varphi(z)$ from (4) the boundary problem results in the following Sturm-Liouville equation:

$$\frac{\partial^2 \varphi}{\partial z^2} - \frac{N^2(z)}{g}\frac{\partial \varphi}{\partial z} + \left(\frac{N^2(z)}{\omega^2}-1\right)\kappa^2 \varphi = 0,$$

(0.6)

$$\frac{\partial \varphi(0)}{\partial z} = g\kappa^2 \varphi(0)/\omega^2, \quad \varphi(-H) = 0,$$

in Boissinesq approximation:

$$\frac{\partial^2 \varphi}{\partial z^2} + \left(\frac{N^2(z)}{\omega^2} - 1\right)\kappa^2 \varphi = 0, \tag{0.7}$$

Where k = | **k** |. Problems (0.6), (0.7) are the problems of the own values, after solution of which, there may be defined the system of the own values $\omega$ (dispersive dependences) and own functions $\varphi(z)$ for each fixed value of the wave number k. The spectrum of such problems is always discrete, that is the system possesses the countable number of the modes $\varphi_n(z)$ (n = 1,2,3...), to each of which there corresponds the law of the dispersion $\omega_n = \omega_n(\kappa)$. In the case when the depth of the liquid is endless and the difference of the function N (z) from zero takes place also within the unlimited interval, then alongside with the discrete spectrum there is also a continuous spectrum.

The knowledge of the dispersive dependences and their properties has the paramount value at research of the linear gravitational waves. The basic properties of the own values and the own functions of the problems (0.6), (0.7) are well studied. The own functions of the considered problems may be divided into to two classes. The first class is presented by one own function $\varphi_0(z)$, which is monotonically and quickly enough decreasing with the increasing depth. This own function poorly depends on the conditions of stratification and describes the superficial wave. All other own functions correspond to the normal modes of the internal waves. For the internal waves own function $\varphi_n(z)$ ( n=1,2,3...) has n-1 zero inside of the interval [-H, 0]. For the continuously stratified liquid of the final depth both for its superficial wave and for its internal waves is typical the monotonous increase in frequency ω of a single mode at the growth of the wave number k, the monotonous reduction of the phase speeds $c_f = \omega/\kappa$ with the growth of k and at the increase of the mode number, and also excess of the phase speeds over the group speeds $c_g = d\omega/d\kappa$. The maximal values of the phase and the group speeds coincide and take place at k = 0. The significant difference of the superficial wave from the internal waves consists that in the short-wave region (κ→∞) the frequency of the superficial wave is unrestrictedly increasing (~ $\kappa^{1/2}$), whereas the internal waves frequency tends to the value $N_M = \max_z N(z)$.

Rather small change of the liquid density at changing the depth in comparison with the drop of the density on the water – air border allows to research the internal waves in the approximation of the "solid cover" ($\varphi(0) = 0$), which filters the superficial waves out without essential distortion of the internal waves. Approximation of "the solid cover" allows to neglect the first sum component in the dynamic condition of (0.2).

The analytical decision of the problems (0.6), (0.7) is possible only for some special cases of changing of *N(z)* function.

At the smooth changing of the function *N(z)* the WKB approximation method is frequently applied for the approximate calculation of the own values and the own functions. However this approach is limited by the case, when the function N(z) has no more than one maximum.

More accurate results may be received by direct use of the numerical methods, and at the present tome there are several methods of the numerical solution of the problems (0.6), (0.7):

1. The finite-difference approximation method, at which the differential equations (0.6), (0.7) and the boundary conditions are replaced with the system of the difference equations.

2. Approximation of the initial continuous distribution of density of the piecewise-constant function. In this case there is a possibility of existence of only the final number of the wave modes. The analysis of the asymptotic behavior of the phases velocities $c_f$ in the shortwave field has demonstrated, that in the liquid with the step-by-step stratification $c_f \approx k^{-1/2}$, while for the liquid with the continuous profile of density $c_f \approx k^{-1}$.

3. Piecewise constant approximation of the Brunt-Väisälä frequency.

4. The numerical solution of the differential equations, derived from (0.6), (0.7) after introduction of the Prewfer modified transformations:

$$\varphi(z) = e^{a(z)} \sin b(z),$$
$$\frac{d\varphi(z)}{dz} = e^{a(z)} \cos b(z)$$

(0.8)

As a result of the transformations (0.8) for definition of the dispersive dependences it is enough to solve the nonlinear boundary value problem of the first order for the function $b(z)$, behavior of which unlike $\varphi(z)$ is monotonous.

The up to now cumulative experience of calculation of dispersive dependences demonstrates, that their most complex behavior arises at the presence in the liquid of several wave guides and on the charts of the dispersive curves there may arise the nodes and crowdings, which testify, that the behavior of the group speeds of the internal waves becomes non-monotonic and on some (abnormal) frequencies the different modes extend practically with the identical phase speeds, having the different group speeds. Such areas are called the resonant zones and in them conditions for an overflow of the energy from the lowest energy-carrying modes into the highest energy-carrying modes are created. This phenomenon looks like as insignificant in application to the linearized problem, but may be important at considering the nonlinear members. The abnormal frequencies represent the rather important feature of the internal waves, on them there is a qualitative change of the vertical structure of the wave field.

The thin structure of distribution of the Brunt-Väisälä frequency also may bring to the similar effects of the crowding of the dispersive characteristics on depth. The dispersive curves under action of the thin hydrological structure can be stratified into the separate groups (clusters) inside which occurs the rapproachement of the dispersive parameters of the different mode, swhereas the groups themselves are moving away from each other. Such stratification, apparently, may affect on the spectra of the internal waves in the field of the frequencies close to the maximum Brunt-Väisälä frequency

For the solution of the equations (0.1) with conditions of (0.2), (0.3) rather convenient method of solution is application of Green functions describing development of generations caused by an instant dot source, being on the depth of $z_1$. In case of the system homogeneous in the horizontal direction it is useful to use Fourier expansion:

$$G(\mathbf{r}, z, z_1, t) = \int \frac{d\mathbf{k}}{(2\pi)^2} \frac{d\omega}{2\pi} e^{i(\mathbf{k}\cdot\mathbf{r} - \omega t)} G(\mathbf{k}, z, z_1, \omega) \qquad (0.9)$$

Then $G(\mathbf{k}, z, z_1, \omega)$ should satisfy the equation of the following kind:

$$\left[ \frac{\partial}{\partial z} \rho_0(z) \frac{\partial}{\partial z} + \rho_0 k^2 \left( \frac{N^2(z)}{\omega^2} - 1 \right) \right] G(\mathbf{k}, z, z_1, \omega) = \delta(z - z_1) \qquad (0.10)$$

The solution of this equation one should look for in the form of the eigenfunctions expansion of the problem (0.6):

$$G(\mathbf{k}, z, z_1, \omega) = G_0(\mathbf{k}, z, z_1) + \sum_n \frac{\omega_n^2(k) \varphi_n(k, z) \varphi_n(k, z_1)}{\omega^2 - \omega_n^2(k)},$$

where $G_0(\mathbf{k}, z, z_1)$ is the solution of the equation (0.10) at $\omega \to \infty$ and describes an instant part of the medium response to the external excitation, and the sum of the eigenfunctions describes the contribution of the wave part. Usually the value $G_0$ is rejected without any discussions. However in some cases, for example, at calculation of the amplitudes of the waves from the periodic sources, this component may be essential, because for the internal waves the law of decrease of the amplitudes of the wave and the non- wave parts of the excitation as the distance from the source of excitation increases is identical.

At fulfillment of the inverse Fourier transformation there is an ambiguity connected with the necessity to set the rule for the flow past the singularities on the real axis $\omega$. The choice of the unambiguous solution is achieved at imposing the causality requirement being reduced to the condition $G(t)|_{t<0} \equiv 0$ (Green's retarded function). Green's retarded function corresponds to the solution satisfying the principle of Mandelshtam radiation, when the energy expands from the source. By virtue of the specific law of dispersion of the internal waves the Mandelshtam radiation condition sometimes does not coincide with the Zommerfeld radiation condition (the waves leaving the source), but the use of the Zommerfeld radiation principle at the choice of the unambiguous solution can lead to the incorrect results.

One more method of the choice of the unambiguous solution is the method attributed to Relay providing for introduction of the infinitesimal dissipation equivalent to the Mandelshtam condition. Often the additional condition is set in the form of the requirement of absence of the wave excitations in the distant area upwards the stream - Long condition, - however the universality of this condition is not obvious at considering the effect of blocking observable in the stratified liquids. It is also possible to use the approach, at which the stationary solution is considered as a limit at $t \to \infty$ of the non-stationary solution for the acting in the stream source of excitation with the constant characteristics, and which is put into operation $t = 0$.

Let us underline, that the causality condition for Green's function is equivalent to the requirement of analyticity of its transient Fourier transformation in the upper half-plane of the complex frequencies $\omega$. It means, that the features on the real axis should be flow past from above, or in accordance with Feynman rule, to exercise the substitution $\omega \to \omega + i\varepsilon$ ($\varepsilon \to +0$), having shifted the features from the real axis downwards. The analyticity of the transient Fourier transformation in the upper half-plane $\omega$ enables to write the Cramers-Cronig ratios expressing relationship between the real and the imaginary parts of Green function, and also in the case of $N = const$ by simple way to construct Green function by means of the analytical continuation from the "non-wave" field of $\omega^2 > N^2$ into the "wave" field of $\omega^2 < N^2$ ( the fields, where the equation of the internal waves belongs accordingly to the elliptic or the hyperbolic type).

The intricate problem is the research of the wave movements in the liquid with the continuous stratification. The majority of the researches conducted in this direction is pertaining to the study of the process of formation of the atmospheric leeward internal gravitational waves at the flow past of the isolated hill or the mountain ridge by the uniform flow. As long as the characteristic speeds and the frequencies of the internal fluctuations are small, then in the most cases it is possible to neglect the influence of the compressibility, viscosity, effects of diffusion and heat conductivity and to use the linear theory of the internal waves. In many researches concerning the leeward waves the authors widely use Long's model. In this model at the special distribution of the speed and the density upwards along the flow the basic equations become linear for any amplitudes of the excitations.

Long's model can be used only at the supercritical modes of the flow past of the obstacle ($\pi V / NH > 1$, where V – the speed of the flow), because at the subcritical currents ($\pi V / NH < 1$) there is the effect directed upwards along the stream, and in this case it is expedient to use other models.

At generation of the flat internal gravitational waves by the bottom irregularity as well as by the superficial pressures and by the dipped source - for the liquid with the various distribution of the density, the structure of the non-fading wave movement behind any excitation is formed from the limited number of the harmonic waves. If to express the maximal phase speeds of the superficial wave and the internal waves through $c_n$ (n = 0,1,2,K), then at $V > c_0$ the continuous waves are absent. At $c_n < V < c_{n+1}$ the generated wave movement is composed out of the exponentially fading trace at $|x| \to \infty$ and the continuous trace formed downwards along the stream and being the sum of the precisely $n+1$ standing harmonic waves.

The Problem of generation of the wave movements at the flow past of the bottom irregularities one can consider as a special case of the more complex problem of generation of the waves at the movement of the submerged body. The final solution represents the sum of three items, the first of which describes the superficial waves fading exponentially at changing of the depth, the second item describes the internal waves caused by the density stratification and existing similarly to the superficial waves but directed downwards along the stream, the third item - the local effects, which are quickly fading with the increase of their distance from the source.

The asymptotic estimation of these items at $\xi \to \infty$ ($\xi = \sqrt{x^2 + z^2}$) shows, that the second item at the big distances represents the sum of two cylindrical waves with their centers in the points $(0,-h)$ and $(0,h)$, which are fading $\approx \xi^{-1/2}$ (where h is the depth of the submerged source).

Simulation of the flow past of the body in the stratified liquid by the simple distribution of the linear features leads to the "paradox" of the infinitely big value of the wave losses in the case of N = const. One of the ways of elimination of this feature is the use of the non-local sources. The other constructive method is introduction of the superficial distribution of the sources, which is borrowed from the theory of the boundless liquid or derived from the solution of the boundary integral equations satisfying to the condition of the flow past of the body by the stream of the stratified liquid. For distributed along the cylinder surface of the energy sources, proceeding from the condition of the flow past, it is possible to obtain the integral equation, which is reduced to the infinite system of the algebraic equations for the coefficients of the forces angular harmonics expansion.

In the plane problem of generation of the internal gravitational waves at the low speeds of movement (Froude number $Fr = NH/V << 1$) there is the blocking phenomenon consisting in the availability of the rather lengthy area of the motionless (blocked) liquid in front of the body, which is flow past by the stratified liquid. The theoretical researches demonstrate, that in the approximation of the very small Froude numbers this phenomenon may be described by the linear equation for the function of the current. The theory claims, that the transversal dimensions of the blocked area are decreasing with increasing distance from the source along the direction of the its movement approximately as $x^{-1}$. This dependence in the sufficient degree is coordinated with the experimental data. In the theory of the leeward waves the discussion of the blocking is basically connected with the so-called Long's condition requiring the absence of excitations caused by the bodies far upwards along the stream.

The rather complex movement of the excitation of the internal gravitational waves is the superposition of the translation movement and the oscillating movement. At such movements depending on the value of the translational speed and the frequency of oscillations there is a reorganization of the wave flow from the symmetric flow (just for the oscillating movement) to the movement existing only downwards along the flow from the excitation (for the uniform translation movement). Availability of the stratification essentially changes the number and the character of the progressive waves, and also leads to formation of two critical speeds $V_{1,2}$ ($V_1 > V_2$), and at that the first speed coincides with the case of the homogeneous liquid and is equal to $V_1 = g/4\sigma$. The complete number of the progressive waves is equal: to eight - at $V < V_2$; to six - at $V_2 < V < V_1$; to four – at $V > V_1$.

At analysis of the system of the ship caused internal gravitational waves the main attention, as a rule, is given to the distant wave zone. As it is known, the accurate linear solution of this problem represents the sum of the double and single integrals, and at the analysis of the distant wave zone one usually confines oneself only to consideration of them. For particular distributions of superficial pressure there may be used the analytical transformation of the double integrals into single integrals, which considerably facilitates the numerical solution and allows to research in details the form of the free surface in the surroundings of the epicenter of the area of the pressures. The gained solutions allow to research the characteristics both of the near, and the distant wave fields. The ratio between the contributions of the different items into the wave field and the values of amplitudes are essentially depended on the speed of the flow and the horizontal dimensions of the obstacle. With reduction of the speed of the stream or at increase of the horizontal dimensions of the obstacle the contribution of the items forming the system of the waves downwards along the flow is decreasing. At that the main excitations are concentrated above the apex of the irregularity of the bottom.

The solution of the linear problem of the flow past of the submerged spheroid by the uniform flow of the liquid of the infinite depth uses the accurate realization of the boundary condition on the contour of the flow past, for which purpose the function of distribution of the sources along the surface of the body is introduced. The presented results of the numerical calculations of the wave resistance demonstrate, that coincidence with the known Havelok approximation, in which the body is simulated by distribution of the sources on the axes (similarly to the theory of the boundless liquid), will be the better, the bigger is the depth of submerging and the less is the lengthening of the spheroid.

The wave field being formed in the liquid consisting of "n"-number of the homogeneous layers is represented, as a rule, in the form of the sum, which is equal to "n" the different wave systems, each of which is in a general way similar to the system of the classical ship waves in the homogeneous liquid of the final depth

At consideration of the problem of formation of the internal gravitational waves by the moving bodies one meets a question of meeting the conditions of the flow past on the surface of the body. Alongside with the attempts to find the exact solution of the flow past problem, for example in the theory of the leeward waves, the wide circulation has the method of substitution of the body by the system of the energy and mass sources simulating the body. In the majority of the publications concerning definition of the excitations caused by the body in the stream stratified in density the body is simulated by the system of the sources and the run-offs corresponding to the solution of the problem of the flow past by the stream of the homogeneous liquid. At that it is intuitively supposed, that at least in the case of the big Froude numbers the similar simulation is rightful. However it was ascertained, that in the spatial problem at simulation of the body horizontally moving in the liquid with $N(z) = const$ system of the dot-type sources the pressure excitation and the number of other hydrodynamic characteristics are turn out to be singular on the axis of the trace. It results in the infinity of the losses of the energy used for radiation of the internal waves and for the wave resistance. Such paradoxes take place for any one-dimensional distributions of the sources along the axis of the horizontal movement. It is possible to get rid of them only at use of the transversal non-localities for simulation of the moving body. The similar singularities are absent in the plane problem for the sources with zero total intensity (dipole), and also at the vertical movement of the sources.

     For elimination of the indicated difficulties it was offered to simulate the body using the system of the sources distributed along its surface, the density of which should be determined from the solution of some boundary integral equation resulting from the requirement to meet the conditions of the flow past on the surface of the body. If, for example, in the capacity of such a equation to use conditions of the flow past of the body by the homogeneous liquid, then it is possible to obtain the final expression for the wave resistance of the sphere. However in the case of Froude small numbers, when characteristic length of the wave is comparable with the dimensions of the body, the interference effects of the internal waves on the surfaces of the body become essential and the approximation of the homogeneous liquid is inapplicable. Accounting of some features of the real flow past of the body (the boundary layer, the wake and etc.) leads to the necessity of simulation of the body using the complex distributions of the sources.

At the fair speeds of movement of the body ($Fr \gg 1$) the wave resistance poorly depends on the concrete kind of the sources. For the horizontal movement of the body the wave resistance at high speeds of the body movement has the universal appearance $A \ln Fr / Fr + B$. The complex wave picture arises at the superposition of the translational movement and the oscillating movement of the sources of excitations. The action of the oscillating superficial pressures "switched on" at the initial moment of the time in the uniform stream of the homogeneous liquid of the final depth is expressed in the integral form. We have in detail studied the case of the infinitely deep liquid and the concentrated pressure $f(x,t) = P_0 \delta(x) \delta(y)$. The study shows, that in this case the solution represents the sum of two items: the first item is proportional to $e^{i\sigma t}$ and describes the steady-state solution (at $r \to \infty$ it decreases $\approx t^{-1}$), the second item is the non-steady solution (at $t \to \infty$ it decreases $\approx t^{-1}$). The detailed study of the wave movement generated by the oscillating area of the superficial pressures is very important at planning and designing of the air-cushion crafts. At that the non-stationary field of pressure can be connected both with the movement of the air-cushion craft itself, and with superficial heavy sea.

 The peculiar phenomenon of the current observable only in the stratified medium is the collapse of the area of mixing. The potential energy (and sometimes and the kinetic energy) in such area is higher, than in the surrounding liquid. In some time under the action of Archimedean forces the area of mixing begins to become flattened in the vertical direction and to be stretched in the horizontal direction working for restoration of the previous state of the stable stratification. This process is accompanied by the intensive generation of the internal gravitational waves in the surrounding liquid.

In the nature formation of such areas of mixing, which frequently are turbulent, is dealt with the various geophysical phenomena. In the problem concerning the collapse the interest is attracted both to the deformation in time of the mixing area itself, and to the system of the internal waves formed in the surrounding liquid. The internal waves considered in the given problem are formed in the time periods comparable with the Brunt-Väisälä period. At that in spite of the fact that the behavior of the liquid in the spot itself is rather intensive, the nonlinear oscillations in the surrounding liquid are rather small and can be described by the linear equations.

The similar to the problems of the collapse of the cylindrical area with the initial generation of the density are the problems of evolution in the stratified liquid of the originally axisymmetric distributions of vorticity (with the horizontal axis of symmetry). The vortical movement causes the infringement of the initial equilibrium distribution of the density in some area, that is accompanied with its collapse.

The problem of the collapse of the area of mixing was subjected to the multiple numerical solution in the various aspects. It is frequently used as a test example at approbation of the new numerical schemes.

One of the interesting questions in the problem of collapse is an exchange of energy between the area of mixing and the surrounding liquid. With the purpose they usually are solving numerically the Navier-Stokes equations in Boissinesq approximation taking into account the diffusion for the linearly stratified liquid at presence of the area of mixing, for which the various initial distributions of the density have been already studied. At the numerical calculation of the currents in the boundless volume there are significant difficulties in the definition of the boundary conditions for the final area of the numerical integration. This difficulty is the consequence of the absence of knowledge of the dynamic behavior of the solution outside the calculation field. One of the methods of solution of this problem is the use of the "open" boundary requirement introduced on the basis of a condition of Zommerfeld radiation, that allows the generated in the calculated area phenomenon to pass through the boundary without some essential distortion and influence on the internal solution..

At the rather intensive collapse of the zone of mixing it generates the single internal gravitational waves with so big amplitude, that on their crests the secondary instabilities may be formed, and the similar kind instabilities apparently may be formed in the ocean at the collapse of the mixed-up areas, and, they, probably, can be registered from the space according to their features on the free surface.

By virtue of the linearity of the problem considered in the given monograph, the field of the internal gravitational waves formed by the non-local source of generations in the stratified liquid is expressed through Green function of the equation of internal waves. In the big number of the research publications there are the exact integral presentations of the function of Green equation for the internal waves, and also the asymptotic presentations of Green functions making it possible to describe both the near and the distant wave fields. There are the methods of calculations of the fields of the internal waves generated by a moving non-local source of generations of the arbitrary shape. Solution of the problem of the field of the internal waves excited by the non-local source represents, as a rule, the sum of the triple quadratures with fast-oscillatory phase functions. The research of the similar sort of the quadratures usually provides for assumption, that the form of the non-local source of the generations is described by a rather simple function (a sphere, a cylinder, an ovoid, etc.), and besides they use the assumption , that the exponential dependence of the density on the depth. In this case it is possible to analyze the gained solutions by the analytical method. Sufficiently many publications are devoted to the exact numerical study of the corresponding linearized equations of hydrodynamics, however for calculation of the distant fields of the internal gravitational waves the usage of only the numerical methods seems inefficient.

The distribution of the internal gravitational waves in the stratified mediums (the ocean, the atmosphere) depends on the horizontal heterogeneity and the nonstationarity of the parameters of the mediums. To the most distinctive horizontal non-uniformities of the real ocean it is possible to consider the relief of its bottom, non-uniformities of the field of density and changeability of the middle currents. The exact analytical solution of the problem of distribution of the internal gravitational waves in the stratified vertically and horizontally non-uniform and non-stationary mediums may be received by the method of separation of the variables only in the event, if the distribution of the density and the form of the bottom are described by the simple enough model functions. When the form of the bottom and stratification are random, it is possible to construct only asymptotic representations of the solution in the near and the distant areas, however for description of the field of the internal waves between these areas it is necessary to have the exact numerical solution of the problem. If, for example, the depth of the ocean and the density varies slowly in comparison with the characteristic length of the internal waves, that often takes place in the real ocean, then at the solution of the problem on distribution of the internal waves above the variable bottom it is possible to use the method of geometrical optics or its upgrade. At use of the asymptotic representations of the wave field located at the big distances from the source of the generation the problem of construction of the uniform asymptotics of the distant field of the internal waves can be solved, in particular, by means of one of the upgrades of the method of geometrical optics - by the "vertical modes - horizontal beams " method, in which it is not supposed to have the slownesses of change of the medium parameters on the vertical coordinate.

The important and rather complex problem is interpretation and analysis of the data of the natural measurements of the internal gravitational waves, at that, in particular, there emerges the problem of separation of the passing wave trains on the background of the casual interferences, and also determination of the characteristics of the wave trains. The most widespread method of the analysis of the data of the natural supervision is the spectral space-time method and its upgrades. There are also other methods using some integral transformations of the data for separation of the wave components on the background of the casual hydrophysical fields.

The presented results are received during the researches under the projects supported by the Russian fund of the basic researches:

№96--01-01120 " Internal gravitational waves in the non-uniform mediums: excitation, propagation , the analysis of the results of the measurements";

№98--05-64606 "Five approaches to the study of the tidal internal waves in the northern part of Pacific ocean";

№99--01-00856 " Critical phenomena at excitation and propagation of the internal gravitational waves in the non-uniform mediums: the theory and the natural observations";

№93--013-17702 "Asymptotic methods in the linear and nonlinear problems of hydrodynamics and gas dynamics";

№96--01-00937 "Singular asymptotic solutions of the linear and nonlinear equations of hydrodynamics and gas dynamics";

The project INTAS 94-2187 "Nonlinear and singular partial differential equations and applications";

The projects of International Science Foundation M3L000, M3L300 "The evaluation of fields excited by oscillating sources moving in stratified fluids and in general dispersive media: the construction of the uniform asymptotic and the creation of the effective computer programs";

and also within the limits of performance of the Federal Target Program "World ocean" and the program of the Ministry of Science of the Russian Federation " Complex researches of the oceans and the seas, the Arctic regions and the Antarctic", the project " Waves in the ocean ".

## 1.1 Basic equations

The system of equations of hydrodynamics for the ideal (non-viscous) liquid being in the homogeneous field of gravitation, looks like:

$$\rho \frac{dU_1}{dt} + \frac{\partial p}{\partial x} = 0; \qquad \rho \frac{dU_2}{dt} + \frac{\partial p}{\partial y} = 0;$$

$$\rho \frac{dW}{dt} + \frac{\partial p}{\partial z} + g\rho = 0 \qquad (1.1.1A)$$

$$\frac{1}{c^2} \frac{dp}{dt} = \frac{\partial \rho}{\partial t} \qquad (1.1.1B)$$

$$\frac{\partial \rho}{\partial t} + div(\rho \mathbf{U}) = 0 \qquad (1.1.1C)$$

Here: $\mathbf{U} = (U_1, U_2, W)$ - velocity vector, $z$- axis is directed upward, $p, \rho$ - pressure and density;

$$\frac{d}{dt} = \frac{\partial}{\partial t} + U_1 \frac{\partial}{\partial x} + U_2 \frac{\partial}{\partial y} + W \frac{\partial}{\partial z}.$$

The equations (1.1.1A) are the equations of motion; (1.1.1B) is the equation of state (in the adiabatic approximation, c is the speed of the sound); (1.1.1C) is the equation of continuity.

If we consider the liquid layer restricted by the free surface and the bottom , then on these boundaries it is necessary to fix the boundary conditions. The requirement to the bottom is simple - it is reduced to the demand of its leakless, that is to the requirement, that the normal to the bottom component of the speed is equal to zero. In the case of the horizontal bottom $z = -H = $ const . This demand is reduced to the null boundary condition for the vertical component of speed

$$W \equiv 0\big|_{z=-H} \qquad (1.1.2)$$

The requirement to the subjected to determination during the solution of the equation system (1.1.1) free surface $z = \zeta(t, x, y)$ is more complex. It includes two requirements – one is kinematic and another is dynamic. The kinematic requirement needs, that the particle of the liquid on the free surface normal to the surface component of the speed coincides with the speed

of the surface movement. This requirement is reduced to the equality

$$W|_{z=\zeta} = \frac{d\zeta}{dt} = \frac{\partial \zeta}{\partial t} + U_1 \frac{\partial \zeta}{\partial x} + U_2 \frac{\partial \zeta}{\partial y} \qquad (1.1.3A)$$

The dynamic requirement needs, that the surface pressure coincides with the air pressure: $p(x,y,\zeta(x,y,t),t) = p_a(x,y,t)$. Further for the sake of simplicity we shall consider, that the air pressure is equal to zero, and the dynamic requirement we shall take in the form of

$$p(x,y,\zeta(x,y,t),t) = 0 \qquad (1.1.3B)$$

We shall linearize the equations (1.1.2) and boundary conditions (1.1.3) in respect to the quiescent state:

$$U_1 = U_2 = W = 0; \quad \rho = \rho_0(z); \quad p = p_0(z) = -g\int_0^z \rho_0(z)dz$$

For this purpose we shall consider, that $p = p_0 + \hat{p}$; $\rho = \rho_0 + \hat{\rho}$; $U_1 = \hat{U}_1$; $U_2 = \hat{U}_2$; $W = \hat{W}$, then in the gained from (1.1.1) - (1.1.3) equations for $\hat{p}\ \hat{\rho}\ \hat{U}_1\ \hat{U}_2, \hat{W}$ we shall leave only the linear members of the equations. As a result we shall have the equations (further the index $\wedge$ is omitted):

$$\rho_0 \frac{\partial U_1}{\partial t} + \frac{\partial p}{\partial x} = 0$$

$$\rho_0 \frac{\partial U_2}{\partial t} + \frac{\partial p}{\partial y} = 0$$

$$\rho_0 \frac{\partial W}{\partial t} + \frac{\partial p}{\partial z} + g\rho = 0$$

(1.1.4)

$$\frac{1}{c^2}\left(\frac{\partial p}{\partial t} - g\rho_0 W\right) = \frac{\partial \rho}{\partial t} + W\frac{\partial \rho_0}{\partial z}$$

$$\rho_0\left(\frac{\partial U_1}{\partial x} + \frac{\partial U_2}{\partial y} + \frac{\partial W}{\partial z}\right) + \frac{\partial \rho}{\partial t} + W\frac{\partial \rho_0}{\partial z} = 0$$

and the boundary conditions at $z = 0, -H$

$$\left.\frac{\partial p}{\partial t} - Wg\rho_0(0) = 0\right|_{z=0} ;$$

$$W = 0\big|_{z=-H} \quad (1.1.5)$$

If from the equation (1.1.4) to exclude unknown members $U_1, U_2, p, \rho$, then we shall have the equation for vertical component of the speed:

$$-\frac{1}{c^2}\frac{\partial^4 W}{\partial t^4} + \frac{\partial^2}{\partial t^2}\left[\Delta W + \frac{\left(\frac{d\rho_0}{dz}\right)}{\rho_0}\frac{\partial W}{\partial z}\right] + N^2 \Delta_h W = 0 \quad (1.1.6)$$

where $N^2 = -\left[\frac{g}{\rho_0}\frac{d\rho_0}{dz} + \frac{g^2}{c^2}\right]$ - a square of Brunt-Väisälä frequency,

$$\Delta_h = \frac{\partial^2}{\partial x^2} + \frac{\partial^2}{\partial y^2}; \qquad \Delta = \Delta_h + \frac{\partial^2}{\partial z^2}.$$

The equations (1.1.4) or the equivalent to them equation (1.1.6) describe both the acoustic waves, and the internal gravitational oscillations. The speed of sound "c" in the ocean water has the order 1.5·10 m\s, the typical speed of the internal waves is no more than 1 m\s, so it is less than the acoustic speed in one and a half thousand times. Therefore the acoustic waves and the internal gravity waves are spreading "non-hindering" each other, and it is not difficult to draw the approximate equations describing either only acoustic waves, or only the internal waves.

To describe only acoustic oscillations, it is necessary to introduce the " fast time" $\tau = ct$, for which the acoustic speed shall have the order of unity. In variables $\tau, x, y, z$ the equation (1.1.6) will be recorded in the following form

$$\frac{\partial^2}{\partial \tau^2}\left[\frac{\partial^2 W}{\partial \tau^2} - \Delta W - \frac{\left(\frac{d\rho_0}{dz}\right)}{\rho_0}\frac{\partial W}{\partial z}\right] + \frac{1}{c^2}N^2 \Delta_h W = 0$$

and, at $c \gg 1$ casting out the last item, we shall gain the wave equation

$$\rho_0 \frac{\partial^2 W}{\partial \tau^2} = \text{div}[\rho_0 \text{grad} W]$$

To filter off the acoustic oscillations, that is to make out the equations describing only the internal waves, it is necessary in the equations (1.1.4) or (1.1.6) to conduct the limiting process at $c \to \infty$, that is to use the approach of the incompressible liquid. Then we shall have:

$$\rho_0 \frac{\partial U_1}{\partial t} + \frac{\partial p}{\partial x} = 0$$

$$\rho_0 \frac{\partial U_2}{\partial t} + \frac{\partial p}{\partial y} = 0$$

$$\rho_0 \frac{\partial W}{\partial t} + \frac{\partial p}{\partial z} + \rho g = 0 \qquad (1.1.7A)$$

$$\frac{\partial U_1}{\partial x} + \frac{\partial U_2}{\partial y} + \frac{\partial W}{\partial z} = 0; \qquad \frac{\partial \rho}{\partial t} + W \frac{\partial \rho_0}{\partial z} = 0 \qquad (1.1.7B)$$

and for the function W - the equation

$$\frac{\partial^2}{\partial t^2} \left[ \Delta W + \frac{\left(\frac{d\rho_0}{dz}\right)}{\rho_0} \frac{\partial W}{\partial z} \right] + N^2 \Delta_h W = 0 \qquad (1.1.8)$$

where $N^2 = -\frac{g}{\rho_0} \frac{d\rho_0}{dz}$

the Equation (1.1.8) is neither the hyperbolic equation, nor the elliptical equation; it falls into to the class of the equations for the first time analyzed by Mr. S.L.Sobolev and now called equations by Sobolev -Galeperin.)    At study of the internal waves in the marine environment frequently for simplification of the calculations it is possible to use one more approximation - Boissinesq approximation.

The density change at ocean makes 3-4 %. Thefore in the equation s (1.1.7A) it is possible to replace $\rho_0(z)$ with the constant value $\rho_0 = \text{const}$ and to consider dependence of $\rho_0$ on z only in the second equation (1.1.7B), supposing $\frac{1}{\rho_0}\frac{\partial \rho_0}{\partial z} = -\frac{N^2}{g}$. Then the system of the equations (1.1.7) will be recorded in the following form:

$$\rho_0 \frac{\partial U_1}{\partial t} + \frac{\partial p}{\partial x} = 0; \quad \rho_0 \frac{\partial U_2}{\partial t} + \frac{\partial p}{\partial y} = 0; \quad \rho_0 \frac{\partial W}{\partial t} + \frac{\partial p}{\partial z} + \rho g = 0 \qquad (1.1.9)$$

$$\frac{\partial U_1}{\partial x} + \frac{\partial U_2}{\partial y} + \frac{\partial W}{\partial z} = 0; \quad \frac{\partial \rho}{\partial t} - W\frac{N^2}{g} = 0$$

and for the function W we shall have the equation

$$\frac{\partial^2}{\partial t^2}\Delta W + N^2(z)\Delta_h W = 0 \quad (1.1.10)$$

Now let's formulate the ratio of the energy for the internal waves. The element of volume dV possesses the kinetic energy

$$\delta E_k = \frac{1}{2}\rho_0(U_1^2 + U_2^2 + W^2)dV$$

Its potential energy caused by the density gradient $\frac{d\rho_0}{dz}$ and the elevation $\zeta(x,y,z,t)$ of the considered volume of the liquid in respect to its neutral (undisturbed) state is equal to

$$\delta E_p = -\frac{1}{2}g\frac{d\rho_0}{dz}\zeta^2 dV = \frac{\rho_0 N^2 \zeta^2}{2}dV$$

But the elevation $\zeta(x,y,z,t)$ and disturbance of density $\rho(x,y,z,t)$ are coupled by the ratio:

$\rho = -\zeta(x,y,z,t)\frac{d\rho_0}{dz}$. Considering, that $\frac{d\rho_0}{dz} = -\frac{\rho_0}{g}N^2$, we shall have the following expression for the potential energy of the concidered volume of the liquid

$$\delta E_p = \frac{1}{2}\frac{g^2\rho^2}{\rho_0 N^2}dV$$

Therefore the density of the total energy of the field of the internal waves is equal to

$$E = \frac{1}{2}\left[\rho_0(U_1^2 + U_2^2 + W^2) + \frac{g^2\rho^2}{\rho_0 N^2}\right]$$

From the equations (1.1.7) follows, that

$$\frac{\partial E}{\partial t} + div(p\mathbf{U}) = 0$$

where $\mathbf{U} = (U_1, U_2, W)$ is the speed vector. Having integrated this expression by some volume V with the boundary S, we shall have

$$\frac{d}{dt}\int_V E dV = \int_S p\mathbf{U}_n dS \qquad (1.1.11)$$

The vector field $p\mathbf{U}$ we naturally may call as the energy flow of the internal waves. The ratio (1.1.11) means, that the increment of the total energy of the internal waves in some volume V is equal to the energy flow inside the boundary S of this volume.

**1.2 Planar internal gravity waves in the exponentially stratified medium**

The most simple view the equations of the internal waves gain in the case of the exponentially stratified liquid, in which the non-disturbed density is the exponential function of the depth: $\rho_0(z) = e^{-qz}$. In such medium $N^2(z) = gq = \text{const}$ and the equation (1.8) becomes with the equation with the constant factors. For simplicity, limiting to Boissinesq approximation, we shall gain from (1.1.10)

$$\frac{\partial^2}{\partial t^2}\left(\frac{\partial^2 W}{\partial x^2} + \frac{\partial^2 W}{\partial y^2} + \frac{\partial^2 W}{\partial z^2}\right) + N^2\left(\frac{\partial^2 W}{\partial x^2} + \frac{\partial^2 W}{\partial y^2}\right) = 0 \qquad (1.2.1)$$

and the system of equations (1.1.9) transforms into

$$\frac{\partial U_1}{\partial t} + \frac{\partial p}{\partial x} = 0; \qquad \frac{\partial U_2}{\partial t} + \frac{\partial p}{\partial y} = 0; \qquad \frac{\partial W}{\partial t} + \frac{\partial p}{\partial z} + \rho g = 0 \qquad (1.2.2)$$

$$\frac{\partial \rho}{\partial t} - \frac{N^2}{g} W = 0; \qquad \frac{\partial U_1}{\partial x} + \frac{\partial U_2}{\partial y} + \frac{\partial W}{\partial z} = 0$$

The elementary solutions of these equations are the planar waves with dependence on $t$, $\mathbf{r} = (x, y, z)$ in the form of $\exp i\varphi = \exp i(-\omega t + \mathbf{kr})$, where $\mathbf{k} = (\alpha, \beta, \gamma)$ - the wave vector.

Lets turn the axes $x, y$ so, that $\beta$ became equal to zero, and the considered field shall not depend on $y$ (then, in particular, $\frac{\partial U_2}{\partial t} = -\frac{\partial p}{\partial y} = 0$). Assuming, that $U_2 = 0$; $U_1 = U_{10}\, e^{i\Phi}$; $W = W_0\, e^{i\Phi}$; $p = p_0\, e^{i\Phi}$; $\rho = \rho_0\, e^{i\Phi}$, we shall receive from (1.2.1) – the disperse equation.

$$\omega^2(\alpha^2 + \gamma^2) = N^2 \alpha^2 \qquad (1.2.3)$$

and from (1.2.2) - the system of equations concerning $U_{10}, W_0, p_0, \rho_0$:

$$\omega U_{10} = \alpha\, p_0$$

$$\omega W_0 = \gamma\, p_0 + ig\rho_0$$

$$\omega \rho_0 = -\frac{iN^2}{g} W_0$$

$$\alpha U_{10} + \gamma W_0 = 0$$

Requirement of the solubility of this system is fulfillment of the dispersion equation (1.2.3). We assume $\alpha = k\cos\theta$; $\gamma = k\sin\theta$; $|\theta| < \dfrac{\pi}{2}$.

Restricting our consideration for distinctness to the waves spreading in the positive direction of an x axis, we shall derive ( U - a constant amplitude factor)

$$\omega = \dfrac{N\alpha}{\sqrt{\alpha^2 + \gamma^2}} = N\cos\theta \quad (1.2.5A)$$

$$U_1 = U\sin\theta \exp i[-\omega t + k(x\cos\theta + z\sin\theta)]$$

$$W = -U\cos\theta \exp i[-\omega t + k(x\cos\theta + z\sin\theta)]$$

$$p = \dfrac{NU}{k}\sin\theta \exp i[-\omega t + k(x\cos\theta + z\sin\theta)] \quad (1.2.5B)$$

$$\rho = \dfrac{iNU}{g}\exp i[-\omega t + k(x\cos\theta + z\sin\theta)]$$

Lets comment on the presented ratios. The oscillation frequency $\omega$ does not depend on the module $k$ of the wave vector and is determined only by its direction and does not exceeds the frequency of buoyancy $N$. The maximum value $\omega$ gains at $\theta = 0$, in this case

$$p = U_1 = 0; \quad W = -Ue^{-iNt+ikx}; \quad \rho = -\dfrac{iNU}{g}e^{-iNt+ikx}$$

that is the wave vector is oriented in the horizontal direction, and the medium experiences only the vertical oscillations. At $\theta \to \pm\dfrac{\pi}{2}$ in extreme case $\omega = 0$, and we gain the stationary motion, in which $W = 0$; $\quad U_1 = Ue^{\mu ikz}$; $\quad \rho = \dfrac{iNU}{g}e^{\mu ikz}$; $\quad p = \pm\dfrac{NU}{k}e^{\mu ikz}$ that is there is a shift horizontal current with the harmonic dependence on $z$ and the density disturbance (with corresponding to it disturbance of pressure) with the same dependence from $z$.

It should be noted, that for the wave equation and the system of Maxwell equations the dispersion equation is given by $\omega = c|k|$ combined, that is $\omega$, on the contrary, does not depend on the direction of the wave vector and is determined only by its length.

From the last equation in (1.2.4) follows, that the speed vector ($U_{10} e^{i\varphi}, U_{20} e^{i\varphi}$) is perpendicular to the wave vector ($\alpha, \beta$), that is that the speed vector is directed along the wave fronts – the lines of the constant phase $\Phi(t, x, z) = \omega t - \alpha x - \gamma z = \text{const}$. Therefore, it might seem, that the internal gravity waves, similarly to Maxwell equations (where for the planar wave the vectors **E** and **H** are also directed along the wave fronts), are transversal waves. However, such a conclusion is premature.

According to the general theory of oscillations in the mediums with dispersion, the vector of the group velocity, that is the traveling velocity of the wave trains with the high-frequency filling up, is determined by the ratio

$$U_{2\alpha} = \frac{\partial \omega(\alpha, \gamma)}{\partial \alpha}; \qquad U_{2\gamma} = \frac{\partial \omega(\alpha, \gamma)}{\partial \gamma} \qquad (1.2.6)$$

where $\omega(\alpha, \gamma)$ are calculated from the dispersion ratio (1.2.5A)

$$\frac{\partial \omega}{\partial \alpha} = \frac{\partial}{\partial \alpha}\left[\frac{N\alpha}{\sqrt{\alpha^2 + \gamma^2}}\right] = \frac{N \sin^2 \theta}{k}; \qquad \frac{\partial \omega}{\partial \gamma} = -\frac{N \sin \theta \cos \theta}{k}$$

We can see, that the direction of the vector of the group speed coincides with the direction of the vector ($U_{10}, W_0$). The same direction has the average value of the energy flux vector of density ($pU_1, pW$) for the period of $T = \frac{2\pi}{\omega}$. At calculation of the ($pU_1, pW$) it is necessary to consider, that the complex functions (1.2.5B) do not have the physical sense, but their real parts have, and

$$U_1 = |U| \sin \theta \cos(-\omega t + \Psi)$$

$$W = -|U| \cos \theta \cos(-\omega t + \Psi)$$

(1.2.7)

$$p = \frac{N|U|}{k} \sin \theta \cos(-\omega t + \Psi)$$

where $\Psi = \arg U + k(x\cos\theta + z\sin\theta)$ and at calculation of the energy flux it is necessary in the capacity of $U_1, W, p$ to take expressions (1.2.7).

To the analogous direction of propagation we shall come, if we shall search for solutions of the equation (1.1.10) by the ray-tracing method, that is to plot for the internal waves the analog of the Wentzel-Kramers-Brillouin (WKB) solution for the wave equation or Maxwell equations.

Таким образом, как минимум в двух отношениях, внутренние волны качественно отличаются от привычных нам акустических и электромагнитных колебаний. Во-первых, частоты колебаний распространяющихся внутренних волн не могут превосходить частоту плавучести $N$. Во-вторых, внутренние волны распространяются не по нормали к волновым фронтам.

Thus, at least in two ratios, internal waves qualitatively differ from customary to us ultrasonic and electromagnetic modes. At first, oscillation frequencies of spread internal waves cannot surpass frequency of floatability $N$. Secondly, internal waves are spread not on a normal to the wave fronts.

## 1.3 Steady-state internal gravity waves and the problem of reflection from the planar boundaries

Let's consider the harmonic oscillations of the exponentially stratified medium, that is the internal waves with the time dependence of the $e^{-i\omega t}$. Assumed that $\rho_0(z) = e^{-qz}$, $N = \sqrt{gq}$ and $W(t,x,y,z) = e^{-i\omega t + \frac{qz}{2}} \Psi(x,y,z)$

we shall gain from (1.1.8)

$$\frac{\partial^2 \Psi}{\partial z^2} - \frac{N^2 - \omega^2}{\omega^2}\left(\frac{\partial^2 \Psi}{\partial x^2} + \frac{\partial^2 \Psi}{\partial y^2}\right) - \frac{q^2}{4}\Psi = 0 \qquad (1.3.1)$$

If $\omega > N$, then this is the elliptical equation; and because the factor at $\Psi$ is less than zero, then it would hold for the principle of the maximum and it cannot describe the propagating waves. The physical meaning have only the steady-state oscillations with frequencies $\omega < N$, when the equation (3.1) is the hyperbolic equation. At that the vertical variable $z$ plays the role of "time". At $q \neq 0$ the equation (1.3.1) is Klein-Gordon equation, at $q = 0$, that is in Boissinesq approximation this is the wave equation. However, although the equation (1.3.1) is the standard equation of the mathematical physics, the boundary problems arising in the internal waves theory for this equation are rather unusual.

For example, if we study the oscillations in a liquid layer $-H < z < 0$, at that the zero boundary condition for the vertical velocity is taken not only at the bottom $z = -H$, but also on the surface $z = 0$ ("approximation of the rigid cover" filtering off the surface waves – see below), then in terms of the classical wave equation it would mean, that we search for the solution identically converted into zero at the values of "time" $z = -H$ and $z = 0$.

Conditions at infinity (the emanation conditions) for the equation (1.3.1) also have a nonstandard form. As well as for the steady-state oscillations equation in the acoustics or the electrodynamics (Helmholtz equation), they are reduced to the requirement, that the emanating fields should carry the energy away at infinity. However, the expressions for the energy flux of the internal waves, as we have seen above, are significantly differ from the analogous expressions for the acoustic oscillations. For the harmonic internal waves the conditions of emanation one may formulate as the requirement, that the field at the grate values of $r = \sqrt{x^2 + y^2 + z^2}$ could be presented as the superposition of the considered above planar waves carrying energy away at infinity.

Let's consider, for example, in Boissinesq approximation (at $q = 0$) Green function of the equation (1.3.1), that is the solution of this equation with $\delta$ is the function $\delta(x, y, z)$ in the right part.

The normal Green function (for the wave equation) looks like

$$G_1 = \begin{cases} \dfrac{c}{\sqrt{(N^2 - \omega^2)z^2 - \omega^2(x^2 + y^2)}} & z > \dfrac{\omega\sqrt{x^2 + y^2}}{\sqrt{N^2 - \omega^2}} \\ 0 & z < \dfrac{\omega\sqrt{x^2 + y^2}}{\sqrt{N^2 - \omega^2}} \end{cases}$$

($c$ -a constant not essential to us). But this function does not satisfy the condition of emanation for the internal waves, and the Green function satisfying to this requirement, looks like

$$G_2 = \begin{cases} \dfrac{c}{2\sqrt{(N^2 - \omega^2)z^2 - \omega^2(x^2 + y^2)}} & z^2 > \dfrac{\omega^2(x^2 + y^2)}{N^2 - \omega^2} \\ \dfrac{ic}{2\sqrt{\omega^2(x^2 + y^2) - (N^2 - \omega^2)z^2}} & z^2 < \dfrac{\omega^2(x^2 + y^2)}{N^2 - \omega^2} \end{cases}$$

Let's now consider for the exponentially stratified medium the problem about reflecting the two-dimensional (not depending on "y") harmonic (with dependence on "t" of the $e^{-i\omega t}$ form) planar waves from the planar interface. For such waves, as we have seen above, "y" is the component of the velocity $U_2$ converted into zero. As $\dfrac{\partial U_1}{\partial x} + \dfrac{\partial W}{\partial z} = 0$, it is possible to introduce the function $\Phi(x, z)$, that

$$U_1 = e^{-i\omega t} \dfrac{\partial \Phi}{\partial z}; \qquad W = -e^{-i\omega t} \dfrac{\partial \Phi}{\partial x} \qquad (1.3.2)$$

then, assuming $\Phi(x,z) = e^{\frac{qz}{2}} \Psi(x,z)$, we shall gain for $\Psi$ the following from (1.3.1) equation

$$\frac{1}{c^2}\frac{\partial^2 \Psi}{\partial z^2} - \frac{\partial^2 \Psi}{\partial x^2} - \chi^2 \Psi = 0 \qquad (1.3.3)$$

where $c = \frac{\sqrt{N^2 - \omega^2}}{\omega}$ and $\chi = \frac{q\omega}{2\sqrt{N^2 - \omega^2}}$

In the terms of the functions $\Phi, \Psi$ it is convenient to formulate the boundary conditions at presence in the liquid of any impermeable surface $S$. On such surface the normal component of velocity is converted into zero, and from (1.3.2) follows, that this component is equal to the longitudinal; derivative $\Phi$. Therefore on $S$ the function $\Phi = \text{const}$ and we may consider, that $\Phi$, and together with it and $\Psi$ are converted on $S$ into zero. The functions $U_1, W, \rho$ are expressed through $\Psi$ in the following formulas

$$U_1 = \exp\left[-i\omega t + \frac{qz}{2}\right]\left(\frac{\partial \Psi}{\partial z} + \frac{q}{2}\Psi\right)$$

$$W = \exp\left[-i\omega t + \frac{qz}{2}\right]\frac{\partial \Psi}{\partial x}$$

(1.3.4)

$$\rho = \frac{-ig}{\omega}\exp\left[-i\omega t - \frac{qz}{2}\right]\frac{\partial \Psi}{\partial x}$$

From the equations (1.1.7) it is obvious, that $p$ may be determined with accuracy up to the constant item. For the derivatives $\frac{\partial p}{\partial x}$ and $\frac{\partial p}{\partial z}$ we have

$$\frac{\partial p}{\partial x} = i\omega e^{-i\omega t - \frac{qz}{2}}\left(\frac{\partial \Psi}{\partial z} + \frac{q}{2}\Psi\right); \qquad \frac{\partial p}{\partial z} = \frac{i(N^2 - \omega^2)}{\omega}e^{-i\omega t - \frac{qz}{2}}\frac{\partial \Psi}{\partial x} \qquad (1.3.5)$$

The equation (1.3.3) ensures compatibility of these expressions.

Thus, for the given planar wave $\Psi_i$ impinging on the planar boundary $S$, it is necessary to find such a departing $S$ planar wave $\Psi_r$, that the sum $\Psi_i + \Psi_r$ was converted on $S$ into zero. To solve this problem first of all it is necessary to indicate the general form of the planar wave satisfying to (1.3.3) and to find the direction of its propagation (certainly not coinciding, as we have already seen, with the direction of its wave vector).

The function $\Psi = e^{i(\alpha x + \gamma z)}$ is the solution of the equation (1.3.3), if $\alpha^2 - c^{-2}\gamma^2 = \chi^2$. On a plane $\alpha, \gamma$ this equation sets a hyperbola with asymptotes $\gamma = \pm c\alpha$. Therefore it is natural to introduce " the hyperbolic angle " $\theta$ and to assume

$$\alpha = \varepsilon\chi \, ch\theta; \quad \gamma = c\chi \, sh\theta; \quad \Psi = e^{i(\alpha x + \gamma z)}$$

(1.3.6)

Here $\varepsilon = \pm 1$ determines the sign of x- of $\alpha$ unit of the wave vector. It is obvious, $|\gamma| < c|\alpha|$, at $\theta \to \pm\infty$ $|\gamma| \to c|\alpha|$, that is the angle between the horizontal and the wave vector increases with increasing of its modulus and at $|\theta| \to \infty$ tends to the value $\varphi = arctg\, c$.

The direction of propagation of the planar wave (1.3.6) is the direction of the energy flux vector. To determine it we should using the formulas (1.3.4), (1.3.5) find $U_1, W, p$, then, having taken real parts of the gained expressions to find the mean values of $pU_1, pW$ for the time period of $T = \dfrac{2\pi}{\omega}$. As a result of it we shall gain vector $\mathbf{\Pi}$:

$$\Pi_x = \dfrac{c\chi\omega}{2} c\varepsilon \, ch\theta; \quad \Pi_z = \dfrac{c\chi\omega}{2}(-sh\theta) \quad (1.3.7)$$

From comparison of formulas (1.3.6) and (1.3.7) it is obvious, that x - units of the wave vector $\mathbf{k} = (\alpha, \beta)$ and the vector of the energy flux $\mathbf{\Pi} = (\Pi_x, \Pi_z)$ have one and the same sign, that is they are directed to one side, and they z - units are directed to the various sides.

It is natural, that the inclination of the planar boundary S should be set by "the hyperbolic angle". We shall consider, that this boundary transits through the origin of the coordinates, and we shall set it by the parametric equation

$$x = q_x \sigma; \qquad z = q_z \sigma \qquad (1.3.8)$$

where $-\infty < \sigma < \infty$. If $|q_x| > c|q_z|$, it is possible to assume $q_x = c\,\text{ch}\theta_S$; $q_z = \text{sh}\theta_S$, that is to set S by the equations

$$x = c\sigma\,\text{ch}\theta_S; \qquad z = \sigma\,\text{sh}\theta_S$$

(1.3.9)

where the inclination S is set by the parameter $\theta_S : -\infty < \theta_S < \infty$. At $\theta_S = 0$ the boundary S is horizontal, therefore the case (1.3.8) we shall term as "almost horizontal boundary".

at $|q_x| < c|q_z|$ it is possible to assume $q_x = c\,\text{sh}\theta_S$; $q_z = \text{ch}\theta_S$, that is to set S by the equation

$$x = c\sigma\,\text{sh}\theta_S; \qquad z = \sigma\,\text{ch}\theta_S \qquad (1.3.10)$$

This case it is natural to term as "the almost vertical boundary". At last, the case $|q_x| = c|q_z|$ we shall term as the critical inclination.

Let's now the wave $\Psi_i$ is set by the formula (1.3.6) at $\varepsilon = \varepsilon_i$; $\theta = \theta_i$, and propagates above the almost horizontal boundary S and impinges on S, that is the energy flux vector $\mathbf{\Pi}$ is directed to S. From (1.3.7) follows, that the last requirement is equivalent to the inequality

$$\varepsilon_i \theta_S + \theta_i > 0 \qquad (1.3.11)$$

The reflected wave we shall search in the form of - $\Psi_r$, where $\Psi_r$ is set by the formulas (3.6) at $\varepsilon = \varepsilon_r$; $\theta = \theta_r$. From the condition of coincidence on S of the of waves phases $\Psi_i$ and $\Psi_r$ we shall gain

$$\varepsilon_i \text{ch}(\theta_S + \varepsilon_i \theta_i) = \varepsilon_r \text{ch}(\theta_S + \varepsilon_r \theta_r)$$

From where:

$$\varepsilon_i = \varepsilon_r;$$

$$\theta_r = -2\varepsilon_i\theta_S - \theta_i \quad (1.3.12)$$

As $\varepsilon_r\theta_S + \theta_r = -(\varepsilon_i\theta_S + \theta_i) < 0$, the wave - $\Psi_r$ spreads in the direction from S.

The task solution about reflection from the almost vertical boundary S is analogously seolved. We shall consider, that the wave (1.3.6) is spread more to the left of S. It is easy to show, that the energy flux vector $\mathbf{\Pi}$ is directed to S at $\varepsilon = 1$ and from S at $\varepsilon = -1$. Therefore the incident wave $\Psi_i$ we shall set by parameter $\varepsilon_i = 1$ and by the parameter $\theta_i$. For the reflection wave $\Psi_r$ we have $\varepsilon_r = -1$, and from the condition of the phases $\Psi_i$ and $\Psi_r$ coincidence on S:

$\sh(\theta_S + \theta_i) = \sh(\theta_r - \theta_S)$, whence

$$\theta_r = \theta_i + 2\theta_S \quad (1.3.13)$$

So, the problem solution about reflection of the planar wave has the various form for the almost horizontal boundary S: $|q_x| > c|q_z|$ and for the almost vertical boundary: $|q_x| < c|q_z|$. In the intermediate case of the critical inclination $|q_x| = c|q_z|$ the considered problem at all has no solution. To the passage to the limit $|q_x| \to c|q_z|$ in formulas (1.3.10) and (1.3.11) corresponds $|\theta_S| \to \infty$, from (1.3.12), (1.3.13) follows, that for the reflection wave $|\theta_r| \to \infty$, that is the length of the wave vector $\mathbf{k}$ of the reflection wave tends to infinity, and at $|q_x| \to c|q_z|$ this wave has no limit.

What is the physical reason for the lack of the solution at the critical inclination of the reflection surface? We consider the problem concerning the harmonic oscillations, that is the steady-state oscillations. However, this problem represents the physical idealization. In the more real formulation we should suppose, that the source of the oscillations is activated in some instant and then harmonically oscillates at the frequency $\omega$. If at $t \to \infty$ the transition process caused by the activation of the source of the oscillations is fading, and the field seeks to the function, oscillating as $e^{-i\omega t}$, then we may speak about the steady-state oscillations. In case of reflection from the boundary with a critical inclination such steady-state oscillations are absent.

## 1.4 Green function of the equation of the internal gravity waves in the exponentially stratified medium.

Section 1.3 has considered the elementary problem of propagation and reflection of the harmonic planar waves in the exponentially stratified medium. Already this problem demonstrated the peculiarity of the internal waves and their difference from the ultrasonic and electromagnetic oscillations. Further we shall consider the problem about propagation of the internal waves excited by the sources with the preset distribution in time and space, that is the problem of solution of the equation

$$LW(t,x,y,z) = f(t,x,y,z) \quad (1.4.1)$$

Here L is the operator of the internal waves, for ease we shall be restricted to Boissinesq approximation, supposing

$$L = \frac{\partial^2}{\partial t^2}\left[\frac{\partial^2}{\partial x^2} + \frac{\partial^2}{\partial y^2} + \frac{\partial^2}{\partial z^2}\right] + N^2(z)\left[\frac{\partial^2}{\partial x^2} + \frac{\partial^2}{\partial y^2}\right]$$

Function $W(t,x,y,z)$ is the vertical velocity; f - the density function of the sources. In case of the pulsing point source of the mass $f = \delta(x)\delta(y)\delta'(z-z_0)\delta''(t)$; for pulsing vertically directed force $f = \delta(t)\delta(z-z_0)(\delta''(x)\delta(y) + \delta''(y)\delta(x))$, etc. In all the cases we shall suppose, that the sources f are put into action at $t \geq 0$, and at $t < 0$ the medium is in the quiescence, that is that

$$W \equiv 0 \quad (t < 0) \quad (1.4.2)$$

If we consider propagation of oscillations into the layer of the finite depth, then it is necessary additionally to introduce the boundary conditions given in the sections 1.1-1.2. By virtue of linearity of thea problem it is enough to consider Green function, that is the solution of $G(t,x,y,z,z_0)$ equation:

$$\frac{\partial^2}{\partial t^2}\Delta G + N^2(z)\left(\frac{\partial^2 G}{\partial x^2} + \frac{\partial^2 G}{\partial y^2}\right) = \delta(x)\delta(y)\delta(z-z_0)\delta(t) \quad (1.4.3)$$

This function has the elementary form for the exponential medium, at $N(z) = \text{const} = N$, that is the equation (1.4.3) has the constant factors. Then the Green function is calculated by Fourier method and at $t > 0$ Fourier method has the form

$$G(t, x, y, z - z_0) = \frac{-1}{2\pi^2 r} \int_{N|\cos\theta|}^{N} \frac{\sin\omega t \, d\omega}{\sqrt{N^2 - \omega^2}\sqrt{\omega^2 - N^2 \cos^2\theta}} \quad (1.4.4)$$

where $r = \sqrt{x^2 + y^2 + (z - z_0)^2}$; $z - z_0 = r\cos\theta$. At $t < 0$ this function according to (1.4.2) is identically converted into zero.

For this formula proving it is necessary to check, at first, that at $t > 0$ Green function G is the solution of a homogeneous equation $LG = 0$ and, secondly, that Green function G satidfies the initial conditions.

$$G\big|_{t=0} = 0; \quad \frac{\partial}{\partial t}\Delta G\bigg|_{t=0} = \delta(x)\delta(y)\delta(z - z_0) \quad (1.4.5)$$

Really, if LG function is converted into zero at $t \neq 0$, then it is concentrated at $t = 0$. And because from (1.4.5) follows, that G is continuous at $t = 0$, and $\frac{\partial \Delta G}{\partial t}$ at $t = 0$ has a breaking, equal to $\delta(x)\delta(y)\delta(z - z_0)$, then making one more differentiation of $\frac{\partial \Delta G}{\partial t}$ by $t$, we shall come to (1.4.3). To proved, that $G(t, x, y, z - z_0)$ at $t > 0$ is the solution of the equation $LG = 0$, we shall record this function in the form of the contour integral

$$G = \frac{1}{8\pi^2} \int_C e^{i\omega t} \frac{d\omega}{\sqrt{\omega^2 - N^2}\sqrt{\omega^2(x^2 + y^2) + (\omega^2 - N^2)z^2}} \quad (1.4.6),$$

where the loop C bypasses the sections of the integration element counterclockwise. It is obvious, that

$$LG = \frac{1}{2\pi} \int_C e^{i\omega t} \Phi(\omega, x, y, z - z_0) d\omega$$

where

$$\Phi(\omega, x, y, z-z_0) = \frac{-1}{4\pi\sqrt{\omega^2 - N^2}} \left[ (\omega^2 - N^2)\left(\frac{\partial^2}{\partial x^2} + \frac{\partial^2}{\partial y^2}\right) + \omega^2 \frac{\partial^2}{\partial z^2} \right] \left[\omega^2(x^2 + y^2) + (\omega^2 - N^2)(z - z_0)^2\right]^{\frac{1}{2}}$$

If at the real $\omega > N$ to make a change of variables: $x = \sqrt{\omega^2 - N^2}\,\xi$;  $y = \sqrt{\omega^2 - N^2}\,\eta$; $z - z_0 = \omega\zeta$, then it will be easily to make sure, that

$$\Phi = \frac{-1}{\omega(\omega^2 - N^2)}\left(\frac{\partial^2}{\partial \xi^2} + \frac{\partial^2}{\partial \eta^2} + \frac{\partial^2}{\partial \zeta^2}\right)\frac{1}{\sqrt{\xi^2 + \eta^2 + \zeta^2}} = \frac{1}{\omega(\omega^2 - N^2)}\delta(\xi)\delta(\eta)\delta(\zeta) = \delta(x)\delta(y)\delta(z - z_0)$$

This formula may be analytically prolonged and in the field of the complex $\omega$; as a result we shall gain

$$LG = \frac{1}{2\pi}\int_C e^{i\omega t}\delta(x)\delta(y)\delta(z - z_0)d\omega = \frac{\delta(x)\delta(y)\delta(z - z_0)}{2\pi}\int_C e^{i\omega t}d\omega = 0$$

Let's check fulfillment of the initial conditions of (1.4.5). The first of these conditions is obvious, the second condition follows from the line of the equalities

$$\lim_{t \to 0} \Delta \frac{\partial G}{\partial t} = \Delta\left[-\frac{1}{2\pi^2 r}\int_{N|\cos\theta|}^{N}\frac{\omega d\omega}{\sqrt{(N^2 - \omega^2)(\omega^2 - N^2 \cos^2\theta)}}\right] = \Delta\left(-\frac{1}{4\pi r}\right) = \delta(x)\delta(y)\delta(z - z_0)$$

Thus, we have proved the formula (1.4.4) for Greene function. From this formula it is possible to draw the conclusion, that $G$ is a superposition of the harmonic oscillations

$$\frac{\sin \omega t}{\sqrt{\omega^2(x^2 + y^2) + (\omega^2 - N^2)(z - z_0)^2}}$$

with the frequencies $\omega$ concluded in the interval $\left(N|\cos\theta| = \frac{N|z - z_0|}{r}, N\right)$.

However we should note, that this statement, despite of its absolutism, is one-sided. With the same degree of reliance it is possible to state, that G - superposition of oscillations with the frequencies ω concluded outside of this interval. Really, the equality takes place

$$G = -\Gamma = \frac{1}{4\pi r}\left[\int_N^\infty \frac{\cos\omega t\, d\omega}{\sqrt{(\omega^2 - N^2)(\omega^2 - N^2\cos^2\theta)}} - \int_0^{N|\cos\theta|} \frac{\cos\omega t\, d\omega}{\sqrt{(N^2 - \omega^2)(N^2\cos^2\theta - \omega^2)}}\right]$$

(1.4.7)

To proved this formula, we shall consider the taken on a direct $Jm\,\omega = \xi$ integral I from the same integrand, that in (1.4.6). Obviously, I does not depend on $\xi$ at $\xi > 0$. Assuming $\xi \to 0$, we shall gain in a limit, that $I = \frac{1}{2}(G + \Gamma)$. On the other hand, assuming, that $\xi \to \infty$, we shall gain, that $I = 0$, whence follows (1.4.7).

Let's discover the asymptotics of Green function at $t \to \infty$ II. As the phase function ωt in the integral (1.4.4) - the linear function ω, this asymptotics is determined by the boundaries of the interval of integration in (1.4.4), at approaching to which the non-exponential function grows accordingly as $(N - \omega)^{-1/2}$ and $(\omega - N|\cos\theta|)^{-1/2}$. Calculating contributions from the points $\omega = N$, $\omega = N|\cos\theta|$ in an asymptotics (1.4.4) by the normal ways of the method of the stationary phase, we shall gain

$$G \approx \frac{-\sin(Nt - \frac{\pi}{4})}{(2\pi N)^{3/2}\sqrt{t}\,r\sin\theta} - \frac{\sin(N|\cos\theta|t + \frac{\pi}{4})}{(2\pi N)^{3/2}\sqrt{t|\cos\theta|}\,r\sin\theta} + O(t^{-3/2}) \qquad (1.4.8)$$

Here the first unit is the standing wave oscillating with Brunt-Väisälä N frequency. The crests of the waves (the constant phase lines) - the wave fronts of the second unit are determined by the ratio

$$|\cos\theta| = \frac{|z - z_0|}{r} = \frac{(n - \frac{1}{4})\pi}{Nt}$$

Each crest of the wave represents the cone with its axis coinciding with the axis $z$; it originates on this axis at $Nt = (n - \frac{1}{4})\pi$, at $t$ increasing the opening of the cone is magnified, and the cone itself is unrestrictedly approaches to the plane of the source. The asymptotics (1.4.8) is inapplicable near to the axis $z$, where the crests of the waves are formed, and near to the plane $z - z_0 = 0$, from which they flow down. This condition is visible directly from the formula (1.4.8) containing in the denominators $\cos\theta$ and $\sin\theta$. Putting it in other way, the asymptotics (1.4.8) is non-uniform.

To gain the asymptotics applicable near to the axis $z$, that is at the value of $|\cos\theta|$ close to unity, we should rewrite the integral (1.4.4) in the following form

$$G = \frac{-1}{2\pi^2 r} \int_{N|\cos\theta|}^{N} \frac{\sin\omega t \, \Phi(\omega) \, d\omega}{\sqrt{N-\omega}\sqrt{\omega - N|\cos\theta|}} \qquad (1.4.9)$$

where

$$\Phi(\omega) = \frac{1}{\sqrt{(N+\omega)(N|\cos\theta|+\omega)}}$$

Assume

$$\Phi(\omega) = A + B\omega + \tilde{\Phi}(\omega)$$

Integral

$$-\frac{1}{2\pi^2 r} \int_{N|\cos\theta|}^{N} \frac{\sin\omega t \, (A + B\omega) \, d\omega}{\sqrt{N-\omega}\sqrt{\omega - N|\cos\theta|}}$$

Using the linear substitution of the variable of integration translating the interval $(N|\cos\theta|, N)$ into (-1,1), we can reduce it to the tabular integrals being expressed through Bessel functions $J_0$ and $J_1$.

If A and B are selected so, that $\tilde{\Phi}(\omega)$ was converted in zero on the boundaries of the interval of integration $N|\cos\theta|$ and $N$, then the second integral

$$-\frac{1}{2\pi^2 r} \int_{N|\cos\theta|}^{N} \frac{\sin\omega t\, \tilde{\Phi}(\omega)\, d\omega}{\sqrt{N-\omega}\sqrt{\omega - N|\cos\theta|}}$$

will seek to zero not slowly, than $t^{-1}$ at any $\cos\theta$. Therefore at $\theta$ close to zero we shall gain

$$G \approx \frac{-1}{8\pi r N \cos\frac{\theta}{2}} \left[ \frac{1+\sqrt{|\cos\theta|}}{\sqrt{|\cos\theta|}} \sin\left(Nt\cos^2\frac{\theta}{2}\right) J_0\left(Nt\sin^2\frac{\theta}{2}\right) - \frac{1-\sqrt{|\cos\theta|}}{\sqrt{|\cos\theta|}} \cos\left(Nt\cos^2\frac{\theta}{2}\right) J_1\left(Nt\sin^2\frac{\theta}{2}\right) \right] + O$$

(1.4.10)

To gain the asymptotics G applicable at the small values of $\cos\theta$, it is necessary to take advantage of the integral representation (1.4.7). The asymptotics of the first integral in (1.4.7) at $t \gg 1$ and the small values of $\cos\theta$ is determined by the contribution of the point $\omega = N$ and calculated by the standard methods. The second integral can be recorded in the following form

$$-\frac{1}{8\pi^2 r} \int_{-N|\cos\theta|}^{N|\cos\theta|} \frac{\cos\omega t\, d\omega}{\sqrt{(N^2 - \omega^2)(N^2\cos^2\theta - \omega^2)}} = J_1 + J_2 =$$

$$= -\frac{1}{8\pi^2 r} \int_{-N|\cos\theta|}^{N|\cos\theta|} \frac{\cos\omega t\, d\omega}{N\sin\theta\sqrt{N^2\cos^2\theta - \omega^2}} + \int_{-N|\cos\theta|}^{N|\cos\theta|} \frac{\cos\omega t\, \Phi(\omega)\, d\omega}{\sqrt{N^2\cos^2\theta - \omega^2}}$$

where

$$\Phi(\omega) = \frac{-1}{8\pi^2 r} \left[ \frac{1}{\sqrt{N^2 - \omega^2}} - \frac{1}{N\sin\theta} \right]$$

is converted into zero at $\omega = \pm N|\cos\theta|$. Therefore $J_2$ seeks to zero at $t \to \infty$ no more slowly, than $t^{-1}$. The integral $J_1$ is expressed through Bessel function, so that at $t \to \infty$ and the small values of $\cos\theta$ we shall gain

$$G \approx \frac{\sin(Nt - \frac{\pi}{4})}{(2\pi N)^{3/2} r\sqrt{t}\sin\theta} - \frac{J_0(Nt\cos\theta)}{8\pi r N \sin\theta} + O(t^{-1}) \quad (1.4.11)$$

Now let's go from Green function to the problem about the point source of mass. As it was already noted above, to the point impulse source of mass placed in the origin of coordinates corresponds the equation for the vertical velocity with the following right part $\delta(x)\delta(y)\delta'(z-z_0)\delta''(t)$, that is

$$W = \frac{\partial^3 G(t,x,y,z-z_0)}{\partial z \partial t^2} \qquad (1.4.12)$$

Restoring for the function W with the help of (1.1.9) the horizontal components of velocity $U_1, U_2$ and the elevation componet $\zeta$ and using the asymptotics (1.4.8) for Green function, we shall gain the following waver field

$$W \approx \frac{N^{3/2}\sqrt{t}\cos\theta\sin\theta|\cos\theta|^{1/2}}{(2\pi)^{3/2}}\cos(Nt|\cos\theta|+\frac{\pi}{4})+O(t^{-1/2}) \qquad (1.4.13A)$$

$$U \approx \frac{N^{3/2}\sqrt{t}\sin^2\theta\sqrt{|\cos\theta|}}{(2\pi)^{3/2}r^2}\cos(Nt|\cos\theta|+\frac{\pi}{4})+O(t^{-1/2}) \qquad (1.4.13B)$$

$$\zeta \approx \frac{\sqrt{Nt}\sin\theta\sqrt{|\cos\theta|}\,\text{sign}(\cos\theta)}{(2\pi)^{3/2}r^2}\sin(Nt|\cos\theta|+\frac{\pi}{4})+O(t^{-1/2}) \qquad (1.4.13C) \text{ Здесь}$$

U -

Here U is the radial component of the horizontal velocity directed from the axis z to the observation point. The azimuth component of the horizontal velocity is identically converted into zero due to the cylindrical symmetry of the problem. For the energy density $\delta E$ of the considered wave field we have

$$\delta E = \delta E_K + \delta E_\Pi = \frac{\rho_0}{2}(U^2+W^2+\zeta^2 N^2) = \frac{tN^3|\cos\theta|\sin^2\theta}{16\pi^3 r^4}+O(1) \qquad (1.4.13E)$$

According to the formulas (1.4.13), the amplitude of the velocity and elevation excited by the $\delta$ - shaped impulse source of mass grows as $\sqrt{t}$; this field with the increase of t become the more and more short-waved, as the wavelength diminishes as $t^{-1}$. The energy density of this field grows as t. At that the full energy - $\iiint \delta E\, dxdydz = \text{const} \iiint \frac{dxdydz}{r^4}$ is equal to infinity, as the given integral is diverging at $r \to 0$.

These as it were paradoxical properties of the considered field are coupled with the fact, that we research the $\delta$ - shaped point source.

The analogous situation occurs and for the wave equation $\dfrac{\partial^2 U_1}{\partial t^2} = c^2 \left( \dfrac{\partial^2 U_1}{\partial x^2} + \dfrac{\partial^2 U_1}{\partial y^2} \right)$, where to Green function

$$G(t,x,y) = \begin{cases} \dfrac{1}{4\pi}(c^2 t^2 - x^2 - y^2)^{-1/2} & ct > \sqrt{x^2 + y^2} \\ 0 & ct < \sqrt{x^2 + y^2} \end{cases}$$

corresponds the continuous integral of energy

$$E = \iint \left\{ \left[ \dfrac{1}{c^2} \dfrac{\partial G}{\partial t} \right]^2 + \left[ \dfrac{\partial G}{\partial x} \right]^2 + \left[ \dfrac{\partial G}{\partial y} \right]^2 \right\} dxdy$$

If we transfer to the source of oscillations spread in space, then these paradoxes will disappear. Let's consider, for example, the internal wave excited by a ball of the radius "a", which suddenly increased it radius by "$\delta a$". To such excitation corresponds the source of the mass uniformly distributed on the surface of the ball of the radius "a", that is the right part in the equation for W equal to $\text{const} \dfrac{\partial^3}{\partial t^2 \partial z} \delta(t) \delta(x^2 + y^2 + (z - z_0)^2 - a^2)$. The field $W_a$ excited by such source of oscillations is expressed through Green function by quadrature

$$W_a = \dfrac{\partial^3}{\partial t^2 \partial z} \iiint G(t, x - \xi, y - \eta, z - z_0 - \zeta) \delta(\xi^2 + \eta^2 + \zeta^2 - a^2) d\xi d\eta d\zeta \qquad (1.4.14),$$

that is it is expressed through the "z" and "t" derivative from the mean value of Green function for the sphere of radius "a". The behavior of the function $W_a$ at the big values of "t" is determined by the ratio of the wavelength of Green function in the point $x, y, z : \lambda = \dfrac{2\pi r}{Nt \sin \theta}$ and the radius of the sphere ".a". At $\dfrac{\lambda}{10} > a$, that is at $t < \dfrac{\pi r}{5 Na \sin \theta}$, Green function practically does not change in the field of integration in (1.4.14), it can be carried out beyond the integral sign, and the function $W_a$ practically does not differ from the function (1.4.13A) corresponding to the point source, that is it grows, as $\sqrt{t}$.

If "t" becomes so big, that $\lambda \ll a$, then the integral (1.4.14) at $t \to \infty$ seeks to zero due to oscillations $G$, and $W_a$ is diminishing like $t^{-1/2}$.

## 1.5 The internal gravity waves in the stratified layer of the finite depth

Further we shall consider propagation of the internal gravity waves in the layer of the finite depth $-H < z < 0$, restricting for ease to Boissinesq approximation. We shall discuss the problem of the boundary conditions. At the bottom, that is at $z = -H$ we put forward the requirement for absence of leakage, that is $W = 0$ ($W$ - a vertical velocity). On the surface at $z = 0$ we had the boundary condition (1.1.5)

$$\frac{\partial p}{\partial t} = Wg\rho_0(0) \qquad (1.5.1)$$

To exclude the pressure $p$ from this requirement, we shall apply to (1.5.1) the functional operator $\Delta_h = \frac{\partial^2}{\partial x^2} + \frac{\partial^2}{\partial y^2}$. From the first two equations (1.1.7A) and the continuity equations (1.1.7B) it follows:

$$\Delta_h p = -\rho_0 \frac{\partial}{\partial t}\left[\frac{\partial U_1}{\partial x} + \frac{\partial U_2}{\partial y}\right] = \rho_0 \frac{\partial^2 W}{\partial t \partial z}$$

Therefore from (1.5.1) it follows, that $\frac{\partial^3 W}{\partial t^2 \partial z} = g\Delta_h W$.

So, in the layer $-H < z < 0$ we consider the equation

$$\frac{\partial^2}{\partial t^2}\Delta W + N^2(z)\Delta_h W = 0 \qquad (1.5.2) \quad \text{with the boundary conditions}$$

$$W = 0 \; (z = -H) \qquad (1.5.3A)$$

$$\frac{\partial^2}{\partial t^2}\left(\frac{\partial W}{\partial z}\right) = g\Delta_h W \; (z = 0) \qquad (1.5.3B)$$

We shall first of all consider the free harmonic oscillations, which have the form of the horizontally running wave

$$W = e^{-i\omega t + i[k_x x + k_y y]} f(z) \qquad (1.5.4)$$

where $(k_x, k_y)$ - the wave vector. Substituting this expression in (1.5.2), (1.5.3), we shall gain the following equations and boundary conditions for $f(z)$ ($k = \sqrt{k_x^2 + k_y^2}$ :

$$\frac{\partial^2 f}{\partial z^2} + \frac{k^2}{\omega^2}(N^2(z) - \omega^2) f = 0 \qquad (1.5.5)$$

$$f = 0 \qquad (z = -H) \qquad (1.5.6A)$$

$$\frac{\partial f}{\partial z} = \frac{k^2 g f}{\omega^2} \qquad (z = 0) \qquad (1.5.6B)$$

The conditions of existence of the nontrivial solution of the boundary problem (1.5.5), (1.5.6) imposes connection (dispersion relations) between the wave number $k$ and the frequency $\omega$. At that it is possible to consider $k$ as the free parameter and to search for eigenfrequencies $\omega = \omega_n(k)$ and eigenfunctions $\varphi_n(z, k)$, that is to consider $\omega$ as spectral parameter; and it is possible to consider $\omega$, on the contrary, as the free parameter and $k$ - as the spectral parameter, and to search for the eigennumber $k_n(\omega)$ and the eigenfunctions $\psi_n(z, \omega)$.

The requirements of orthogonality of the eigenfunctions at $n \neq m$ are defined in the regular way and have the following form

$$g\varphi_n(0,k)\varphi_m(0,k) + \int_{-H}^{0} N^2(z)\varphi_n(z,k)\varphi_m(z,k)\,dz = 0 \tag{1.5.7A}$$

$$g\psi_n(0,\omega)\psi_m(0,\omega) + \int_{-H}^{0}(N^2(z)-\omega^2)\psi_n(0,\omega)\psi_m(0,\omega)\,dz = 0 \tag{1.5.7B}$$

It is obvious, that the functions $\omega = \omega_n(k)$ are the inverse functions to $k = k_n(\omega)$ and that if $\bar{k}$ and $\bar{\omega}$ are coupled by the dispersion relation $\bar{k} = k_n(\bar{\omega})$; $\bar{\omega} = \omega_n(\bar{k})$, then the functions $\varphi_n(z,\bar{k})$ and $\psi_n(z,\bar{\omega})$ coincide with the accuracy up to the constant factor with the solution of the equation (1.5.5) with the boundary condition (1.5.6A) at $k = \bar{k}$; $\omega = \bar{\omega}$, and consequently coincide with each other. This fact does not contradict to the fact, that the ratios of the orthogonality for $\varphi_n(z,k)$ and $\psi_n(z,\omega)$ have the different form, since in (1.5.7A) we take two functions f at one and the same value of $k$, but with different ratios $\omega$: $\omega = \omega_n(k)$ and $\omega = \omega_m(k)$, and in (1.5.7B) we take functions $f(z,k,\omega)$ at one and the same value of $\omega$, but with various values of $k$: $k = k_n(\omega)$ and $k = k_m(\omega)$.

For the equations (1.5.5) and (1.5.6) it is possible to formulate and the other spectral problems. It is possible, for example, to take $\omega = ck$ (where "c" is the phase velocity) and to consider "c" as the free parameter, and "k" - as the spectral parameter. Then the equation (1.5.5), the boundary condition (1.5.6B) and the orthogonality condition will be recorded as follows:

$$\frac{\partial^2 f}{\partial z^2} + \left[\frac{N^2(z)}{c^2} - k^2\right]f = 0 \quad \frac{\partial f}{\partial z} = \frac{gf}{c^2} \qquad (z=0)$$

$$\int_{-H}^{0} f_n(z,c)f_m(z,c)\,dz = 0 \qquad (n \neq m)$$

Selection in (1.5.5) of free and spectral parameters depends on the specific target and the method for its solution. It will be convenient to us to consider further $k$ as the free parameter, and $\omega$ - ae the spectral parameter.

Let's consider the elementary example of a such spectroscopic problem: the layer $-H < z < 0$ with the constant $N(z) = N$. If to remove the values of the parameters, assuming $\xi = \frac{\omega}{N}$, $q = kN$ and to introduce the independent variable $\eta = \frac{z}{H}$, the problems (1.5.5), (1.5.6) will become

$$\frac{\partial^2 \varphi}{\partial \eta^2} + q^2\left(\frac{1}{\xi^2} - 1\right)\varphi = 0 \qquad (-1 < z < 0) \tag{1.5.8}$$

$$\varphi(-1) = 0; \qquad \varphi(0) = \frac{\alpha\xi^2}{q^2}\frac{\partial\varphi(0)}{\partial\eta},$$

where $\alpha = \frac{HN^2}{g} = \left|\frac{H}{\rho_0}\frac{d\rho_0}{dz}\right|$. As in Boissinesq approximation $\frac{1}{\rho_0}\frac{d\rho_0}{dz} = \frac{N^2}{g} = \text{const}$ then

$\alpha = \left|\frac{H}{\rho_0}\frac{d\rho_0}{dz}\right| = \frac{\Delta\rho_0}{\rho_0}$, where $\frac{\Delta\rho_0}{\rho_0}$ - the relative difference of density $\rho_0(z)$ in the whole layer $-H < z < 0$.

Let's consider the key properties of the dispersion functions $\omega_n(k)$ and the corresponding eigenfunctions $\varphi_n(z,k) = f(z,k,\omega_n(k))$ for the case of the arbitrary value of $N(z) > 0$. The eigennumber $\omega_0(k)$, which essentially exceeds all the rest $\omega_n(k)$, grows at accretion of k as $\sqrt{kg}$, weakly depends on N(z) and at the small $N_m = \max_z N(z)$ is close to $\sqrt{kg\,\text{th}\,kH}$. The corresponding eigenfunction $\varphi_0(z,k) = f(z,k,\omega_0(k))$ achieves the maximum at $z = 0$ and promptly diminishes (at the big values of k, as $e^{-k|z|}$) with accretion of the depth $|z|$. This eigenoscillation corresponds to the surface wave.

All the rest eigenfunctions $\varphi_n(z,k)$ (where "n" is the number of reversals of the sign of the function $\varphi_n(z,k)$, that is the number of zeros of this function in the interval $(-H,0)$) achieves its maximum inside the layer $-H < z < 0$. With "n" increase - the dispersion functions $\omega_n(k)$ are diminishing, with the increase of "k" – they are increasing, but always remain less, than $N_m = \max_z N(z)$. The phase velocity $c_f^{(n)} = \omega_n/k$ diminishes with the increase of "k", the complex velocity $c_g^{(n)} = \partial \omega_n / \partial k$ (that is inclination of the dispersion curves) achieves its maximum at "$k = 0$ and tends to zero at $k \to \infty$. The property of the monotonous decreasing of $c_g^{(n)}(k)$ is not kept safe for the intricate profiles N(z). Already for the piecewise constant functions N(z) at the rather great numbers of "n" the dispersion curves $\omega_n(k)$ become non-convex, and the group velocities $c_g^{(n)}(k)$ have the local extremes. Since $\omega_n(k)$ are rather small at $n \geq 1$, in the boundary condition (1.5.6B) the factor at f(0) is great, and so f(0) is small as compared with f'(0). It appears possible, not making a big error to substitute the requirement (1.5.6 B) or correspondingly the requirement (1.5.3 B) for the more simple one (approximation of the solid cover) $W = 0 \quad (z = 0)$.

For example, for the layer with N = const at such substitution the dispersion functions $\xi_n(q)$ at $n \geq 1$ will vary of no more, than by 1 %. At the same time the approximation of the solid cover slightly simplifies calculations and filters the surface waves off. Therefore the approximation of the solid cover is in general usage at study of propagation of the internal waves in the marine medium restricted by the free surface. In the cases, when the rather small, but nevertheless not equal to zero value of the field $W(t,x,y,z)$ represents the autonomous interest, then it is expedient to search at first the field W in the approximation of the solid cover, and then using the obtained values of $\dfrac{\partial W}{\partial z}$ to restore the field W according to the corresponding formulas (1.5.6B), (1.5.3 B).

We should note, at last, that if $N(z) > 0$ at all values of $z: -H < z < 0$, then the system of functions $\varphi_n(z,k)$ is complete.

## 1.6 The internal gravity waves in the stratified mediums with the average shift flows

In the real oceanic conditions often it is necessary to consider the internal gravity waves propagating against the background of the average flows with the vertical shift of velocity. At that the variation of the velocity in the vertical direction makes dozens of cm\s and m\s, that is the same order, as for the maximal velocities of the internal waves. It is clear, that such flows should essentially affect the propagation of the internal waves. If the scale of the flow variations in the horizontal direction is much bigger than the lengths of the internal waves, and the scale of the variations in time is much more than the periods of the internal waves, then the natural model is the case of the stationary and horizontally uniform average flows. Assume $\mathbf{V}(z) = (V_1(z), V_2(z))$ is the flow on the horizon $z$. Linearizing the equations of hydrodynamics concerning the undisturbed state, in which

$$U_1 = V_1(z); \quad U_2 = V_2(z); \quad W = 0; \quad \rho = \rho_0(z); \quad p = p_0(z) = -\int \rho_0 dz$$

And then assuming, that the speed of the sound $c \to \infty$, we shall come to the system of equations, differing from (1.1.4) by the fact, that the operator $\partial/\partial t$ should be substituted for $\dfrac{D}{Dt} = \dfrac{\partial}{\partial t} + V_1(z)\dfrac{\partial}{\partial x} + V_2(z)\dfrac{\partial}{\partial y}$:

$$\rho_0 \frac{DU_1}{Dt} + \frac{\partial p}{\partial x} = 0; \quad \rho_0 \frac{DU_2}{Dt} + \frac{\partial p}{\partial y} = 0; \quad \rho_0 \frac{DW}{Dt} + \frac{\partial p}{\partial z} + \rho g = 0 \qquad (1.6.1)$$

$$\frac{\partial U_1}{\partial x} + \frac{\partial U_2}{\partial y} + \frac{\partial W}{\partial z} = 0;$$

$$\frac{\partial \rho}{\partial t} + W \frac{\partial \rho_0}{\partial z} = 0.$$

Having taken advantage of Boissinesq approximation, that is, assuming here $\dfrac{\partial \rho_0}{\partial z} = -\dfrac{N^2}{g}$ and $\rho_0 = 1$ and then excluding functions $U_1, U_2, p, \rho$, we shall gain the equation for the vertical velocity

$$LW = 0, \qquad (1.6.2)$$

where

$$L = \frac{D^2}{Dt^2}\Delta - \frac{D}{Dt}\left(\frac{\partial^2 V_1(z)}{\partial z^2}\frac{\partial}{\partial x} + \frac{\partial^2 V_2(z)}{\partial z^2}\frac{\partial}{\partial y}\right) + N^2(z)\Delta_h$$

The boundary conditions for (1.6.2) we shall take in the approximation of the solid cover

$$W = 0 \quad (z = 0, -H) \tag{1.6.3}$$

First of all we shall consider the planar waves, that is the solutions of the following form

$$W = e^{i(\lambda x + \mu y - \omega t)} \varphi(z) \tag{1.6.4}$$

Substituting this expression in (1.6.2), we shall gain the equation for $\varphi$ (Taylor-Golshtain equation) and the boundary problem

$$L_0 \varphi = (\omega - kF)^2 \frac{\partial^2 \varphi}{\partial z^2} + \left\{ k^2 \left[ N^2 - (\omega - kF)^2 \right] + k \frac{\partial^2 F}{\partial z^2} (\omega - kF) \right\} \varphi = 0 \tag{1.6.5}$$

$$\varphi = 0 \quad (z = 0, -H)$$

Here we have assumed $\lambda = k \cos \psi$; $\mu = k \sin \psi$; $F = F(z) = V_1(z) \cos \psi + V_2(z) \sin \psi$ - the component of the flow velocity $\mathbf{V}(z)$ in the wave direction (1.6.4).

Let's consider some properties of the spectral problem (1.6.5), where the spectral parameter is $\omega$. Later on it will be convenient for us by substitution of the unknown function

$$\varphi(z) = (\omega - kF(z)) u(z) \tag{1.6.6}$$

to pass to the problem

$$L_1 u = \frac{d}{dz}\left[(\omega - kF)^2 \frac{\partial u}{\partial z}\right] + k^2 \left[N^2 - (\omega - kF)^2\right] u = 0 \tag{1.6.7}$$

In the presence of the average shift flows the internal waves interacting with these flows can interchange of power with them, that is the natural oscillations (1.6.4) can be exponentially damping at ($Jm \omega < 0$) or rising at ($Jm \omega > 0$). With the purpose to avoid such event it is necessary, that the vertical gradient of the average flows was small as compared with the frequency of buoyancy. It is enough to demand, that Miles stability condition, that is that at all values of $z$ the following inequality was fulfilled

$$\left(\frac{\partial V_1(z)}{\partial z}\right)^2 + \left(\frac{\partial V_2(z)}{\partial z}\right)^2 \leq 4N^2(z) \tag{1.6.8}$$

Let's proved, that if the expression (1.6.8) is realized, then the spectral problem (1.6.7) cannot have the complex eigennumbers $\omega = \omega_r + i\omega_i$. For this purpose we shall note, that for any functions $f(z)$ and $g(z)$ converted into zero at $z = 0, -H$, the following equality takes place

$$\int_{-H}^{0} gL_1 f dz = \int_{-H}^{0} k^2 (N^2 - \Phi^2) fg dz - \int_{-H}^{0} \Phi^2 \frac{\partial f}{\partial z} \frac{\partial g}{\partial t} dz \qquad (1.6.9),$$

where

$$\Phi = \omega - kF(z)$$

Let's now consider the case, when $f(z)$ is the solution of the boundary problem (1.6.7) at $\omega = \omega_r + i\omega_i$. Assume $Q = \sqrt{\Phi} f(z)$ and $g = \Phi^{-1/2} Q^*(z)$, where $Q^*(z)$ is the function complexly conjugate with $Q(z)$. Then expression (1.6.9) converts into zero, and its right part can be recorded in the following form

$$\int k^2 \left[ \frac{N^2}{\Phi} - \Phi \right] |Q|^2 dz - \frac{1}{4} \int \frac{1}{\Phi} \left( \frac{\partial \Phi}{\partial z} \right)^2 |Q|^2 dz + \frac{1}{2} \int \left( \frac{\partial \Phi}{\partial z} \right) \frac{d}{dz} (QQ^*) dz - \int \Phi \left| \frac{dQ}{dz} \right|^2 dz$$

The imaginary part of this expression is equal to

$$-\omega_i \left[ k^2 \int \left[ N^2(z) - \frac{1}{4} \left( \frac{\partial F}{\partial z} \right)^2 \right] |\Phi|^{-2} dz + \int \left( k^2 |Q|^2 + \left| \frac{\partial Q}{\partial z} \right|^2 \right) dz \right]$$

since $\left( \frac{\partial F}{\partial z} \right)^2 = \left( \frac{\partial V_1(z)}{\partial z} \cos \psi + \frac{\partial V_2(z)}{\partial z} \sin \psi \right)^2 \leq \left( \frac{\partial V_1(z)}{\partial z} \right)^2 + \left( \frac{\partial V_2(z)}{\partial z} \right)^2 \leq 4N^2(z)$

The expression located in the square brackets, is positive, and consequently it is necessary, that $\omega_i = 0$. Thus, if Miles stability condition (1.6.8) is met, then the spectral problem (1.6.5) (or (1.6.7)) has no complex eigennumbers $\omega$. It is possible to demonstrate, that there are two families of the real eigennumbers $\omega$. In the first of them we shall $\omega_n$ are increasing and seek to $kF_+ = \min_z kF(z)$, in the second family $\omega_n$ are decreasing and seek to $kF_+ = \max_z kF(z)$. The first family we shall number by the negative values $n$, the second family – by the positive values; $|n|$ is the number of the reversals of sign of the eigenfunction $\varphi_n$.

We shall mark the qualitative difference between the behavior of the eigenfunctions $\varphi_n(z)$ at $|n| \to \infty$ for both in the absence or presence of the flows. If the flows are absent, the equation for $\varphi_n$ are looking like

$$\omega^2 \frac{d^2\varphi}{dz^2} + k^2(N^2 - \omega^2)\varphi = 0$$

$\omega_n \to 0$ at $n \to \infty$, and eigenfunctions $\varphi_n(z)$ become more and more oscillating and do not seek at any limit. At the same time at presence of the flows and, for example at $n \to \infty$, $\omega_n \to kF_+ = \max_z kF(z)$, then the equation (1.6.7) seeks to the limiting equation

$$\frac{d}{dz}\left[(F_+ - F(z))^2 \frac{\partial u}{\partial z}\right] + \left[N^2(z) - k^2(F_+ - F(z))^2\right]u = 0 \qquad (1.6.10)$$

and the eigenfunctions $u_n(z)$ at any fixed $z$, for which $F(z) \neq F_+$, and $n \to \infty$ are seeking to the solution of $u(z)$ of this equation.

Green function at presence of the average shift flows satisfies the equation

$$LG(t, r, z, z_0) = \delta(t)\delta(z - z_0)\delta(x)\delta(y) \qquad (1.6.11$$

где L - оператор (1.6.2), нулевым граничным условиям (1.6.3) и начальным условиям

$$G \equiv 0 \quad (t < 0) \qquad (1.6.12)$$

where L is the operator of (1.6.2), satisfies the zero boundary conditions (1.6.3) and the initial conditions

$$G \equiv 0 \quad (t < 0) \qquad (1.6.12).$$

As well as in the case of absence of the flows, using Fourier method, we shall gain

$$G = \frac{1}{(2\pi)^3} \int_{-\infty}^{\infty}\int_{-\infty}^{\infty} e^{i(\lambda x + \mu y)} d\lambda d\mu \int_{-\infty+i\varepsilon}^{\infty+i\varepsilon} e^{-i\omega t} \varphi(\omega, \lambda, \mu, z, z_0) d\omega \qquad (1.6.13)$$

where $\varphi$ is the solution of the equation and the boundary problem

$$L_0\varphi = -\delta(z - z_0); \quad \varphi = 0 \quad (z = 0, -H) \qquad (1.6.14)$$

and $L_0$ is the operator of Taylor-Golfstain (1.6.5). At $Jm\,\omega \neq 0$ there is a unique value of the solution of this problem, as $L_0$ has no complex eigenvalues.

In the case of absence of the flows, we shall solve the equation analogous to (1.6.14), expanding $\delta$-function into series of eigenfunctions of the operator $L = \omega^2 \frac{\partial^2}{\partial z^2} + k^2(N^2 - \omega^2)$ (for the details see Chapter 2). But at presence of the flows the eigenfunctions are not only non-orthogonal, but also they are not complete, and for construction of the solution of (1.6.14) it is necessary to use other method.

Let's designate through $v_1(z,\omega)$ and $v_2(z,\omega)$ the solutions of the equation $L_0 v = 0$ converted into zero accordingly at $z = 0$ and $z = -H$. Then the solution $\varphi(\omega, \lambda, \mu, z, z_0)$ of the boundary problem (1.6.14) shall have the form

$$\varphi(\omega, \lambda, \mu, z, z_0) = \begin{cases} -\dfrac{v_1(z,\omega)v_2(z_0,\omega)}{(\omega - F(z_0))^2 W} & z > z_0 \\ -\dfrac{v_1(z_0,\omega)v_2(z,\omega)}{(\omega - F(z_0))^2 W} & z < z_0 \end{cases} \quad (1.6.15)$$

where: $W = W(\omega) = \dfrac{\partial v_1}{\partial z} v_2 - v_1 \dfrac{\partial v_2}{\partial z}$ are wronskian functions $v_1(z,\omega)$ and $v_2(z,\omega)$.

Now we shall consider the behavior of $v_1$, $v_2$, wronskian $W$ and $\varphi(\omega, \lambda, \mu, z, z_0)$, as the functions of $\omega$. If $\mathrm{Jm}\,\omega \neq 0$, the factor at $\dfrac{\partial^2 \varphi}{\partial z^2}$ in (1.6.5) at any value of $z$ is not converted into zero and consequently the solutions of $v_1$ and $v_2$ in this equation are regular at all values of $z$ and are analytic functions of $\omega$. The values of $\omega$, at which $W$ is converted into zero, are eigenvalues of the operator $L_0$. As Miles stability condition is supposed to be fulfilled, this operator has no complex eigenvalues, $W$ has no complex zero and $\varphi$ is analytical at all complex values of $\omega$.

If $\omega$ is real, but $\omega < kF_- = k \min_z F(z)$ or $\omega > kF_+ = k \max_z F(z)$, then the factor at $\dfrac{\partial^2 \varphi}{\partial z^2}$ in (1.6.5) as before at no values of $z$ converts into zero and the functions $v_1$, $v_2$ are the analytic functions of $\omega$. But the wronskian W can already be converted into zero; its zeroes are eigenvalues of $\omega_n$. The residues of $\varphi$ at $\omega = \omega_n$ are expressed through $\left.\frac{\partial W}{\partial \omega}\right|_{\omega=\omega_n}$, that is through

$$d_n = \left.\frac{\partial}{\partial \omega}\left[\frac{\partial v_2(-H)}{\partial z} v_1(-H)\right]\right|_{\omega=\omega_n} = \left.\frac{\partial \varphi_n}{\partial \omega}\frac{\partial \varphi_n}{\partial z}\right|_{z=-H}$$

For $d_n$ the following valid expression is valid

$$d_n = 2\int_{-H}^{0} (\omega_n - kF)\left\{\left[\frac{d}{dz}\frac{\varphi_n}{(\omega_n - kF)}\right]^2 + \frac{k^2\varphi_n^2}{(\omega_n - kF)^2}\right\}dz \qquad (1.6.16)$$

At last, at $kF_- < \omega < kF_+$ there are values of $z$, at which the factor $(\omega - kF)^2$ at $\frac{\partial^2\varphi}{\partial z^2}$ in (1.6.5) is converted into zero.

This value of $z$ is the point of branching for solution of the equation (1.6.5). Therefore the interval $kF_- < \omega < kF_+$ is a section for the functions $v_1$, $v_2$ and the wronskian W. If $\omega$ lies inside this interval, then the limits of $\varphi(\omega + i\varepsilon)$ and $\varphi(\omega - i\varepsilon)$ (where $\varepsilon \to 0$, and $\varphi$ has the form of (1.6.15) are complexly conjugated and differing from each other. At that there is the possibility to proved, that W is not converted into zero on the upper or the lower edges of this section, that is that all zeroes of W are limited by the indicated above series of $\omega_{-n}$ and $\omega_n$.

Now we may calculate the integral by $\omega$ in (1.6.13). At $t < 0$, carrying the path of integration at infinity in the upper half plane, we shall gain zero. At $t > 0$ the path of integration is enclosed in the lower half-plane, and the integral is reduced to the total of residues and the integral on the cross-section

$$\Gamma = \int_{-\infty+i\varepsilon}^{\infty+i\varepsilon} \varphi e^{-i\omega t} d\omega = 2\pi i \sum_{n=-\infty}^{\infty} e^{-i\omega_n t} \frac{\varphi_n(z)\varphi_n(z_0)}{d_n(\omega_n - kF(z_0))^2} + \Gamma_m$$

(1.6.17)

Here the summation is conducted using eigenfunctions of the operator (1.6.5), that is on the discrete spectrum, and $\Gamma_m$ - the integral on the cross-section, that is on the continuous spectrum of this operator:

$$\Gamma_m = 2i \int\limits_{kF_-}^{kF_+} e^{-i\omega t} Jm\, \varphi(\omega + i0, \lambda, \mu, z, z_0)\, d\omega$$

At the absence of the flows the formula is analogous to (1.6.17) and has the following form:

$$g = \frac{1}{4\pi^2} \sum_{n=-\infty}^{\infty} \frac{1}{2ik^2} e^{-i\omega_n t} \omega_n(k)\, \varphi_n(z,k)\, \varphi_n(z_0,k) \qquad (1.6.18)$$

where we at $n < 0$ have assumed $\omega_n(k) = -\omega_{-n}(k)$; $\varphi_n(z,k) = \varphi_{-n}(z,k)$.

The qualitative difference of the formulas (1.6.17) and (1.6.18), besides the presence in (1.6.17) of integrals on the continuous spectrum $\Gamma_m$, consists in the nature (character) of the convergence of the discrete spectrum. In the formula (1.6.18) the eigenvalues $\omega_n$ behave as $n^{-1}$, and the series (1.6.18) converges conditionally, due to oscillations $\varphi_n(z,k)\,\varphi_n(z_0,k)$. In the case of presence of the flows and, for instance $n \to \infty$, $\omega_n$ are seeking to $-kF_+$, the function $\varphi_n(z)$ and $\varphi_n(z_0)$ seek to the solutions of the equation (1.6.10), and the convergence of the series (1.6.17) is ensured due to the quick growth of $d_n$. It is possible to show, that $d_n \approx n^3$ at $n \to \infty$, that is the series (1.6.17) converges absolutely.

We shall note, by the way, that from that condition, that the decomposition (1.6.17) of the functions $\Gamma$ includes the integral on the continuous spectrum, follows, that this decomposition is short of a series of eigenfunctions, that is the system of eigenfunctions $\varphi_n(z)$ is incomplete.

Eigenfunctions $\varphi_n(z)$ in (1.6.17) depend on $\lambda, \mu$. It is easy to see, that

$$\varphi_{-n}(z,-\lambda,-\mu) = \varphi_n(z,\lambda,\mu); \qquad \omega_{-n}(-\lambda,-\mu) = -\omega_n(\lambda,\mu)$$

Considering these ratios and having conducted integration (1.6.17) by $\lambda$ and $\mu$, we shall gain the following expression for Green function

$$G = \sum_{n=1}^{\infty} G_n(t,x,y,z,) + G_m(t,x,y,z,)$$

Where: $G_n = \dfrac{1}{2\pi^2} Jm \int\limits_{-\infty}^{\infty} \int\limits_{-\infty}^{\infty} e^{i(\lambda x + \mu y - \omega_n t)} \dfrac{\varphi_n(z)\varphi_n(z_0)\,d\lambda d\mu}{d_n(\omega_n - kF(z_0))^2}$ is the n-th mode and

$$G_m = \frac{1}{(2\pi)^2} \iint e^{i(\lambda x + \mu y)} \Gamma_m(t, \lambda, \mu, z, z_0) d\lambda d\mu$$ - is the integral on the continuous spectrum.